\documentclass{jfm}

\usepackage{graphicx}

\usepackage{newtxtext}
\usepackage{newtxmath}
\usepackage{natbib}
\usepackage[normalem]{ulem}
\usepackage{hyperref}
\usepackage{tikz}
\hypersetup{
	colorlinks = true,
	urlcolor   = blue,
	citecolor  = black,
}

\newcommand{\bm}[1]{\boldsymbol{#1}}

\DeclareMathOperator*{\argmin}{argmin} 

\title{Multi-scale data reconstruction of turbulent rotating flows with Gappy POD, Extended POD and Generative Adversarial Networks}

\author{
	Tianyi Li\aff{1,2},
	Michele Buzzicotti\aff{1},
	Luca Biferale\aff{1}
	\corresp{\email{biferale@roma2.infn.it}},
	Fabio Bonaccorso\aff{1},
	Shiyi Chen\aff{2,3}
	\and Minping Wan\aff{2,3}
}

\affiliation{
	\aff{1}Department of Physics and INFN, University of Rome ``Tor Vergata", Via della Ricerca Scientifica 1, 00133, Rome, Italy
	\aff{2}Guangdong Provincial Key Laboratory of Turbulence Research and Applications, Department of Mechanics and Aerospace Engineering, Southern University of Science and Technology, Shenzhen, Guangdong 518055, PR China
	\aff{3}Guangdong-Hong Kong-Macao Joint Laboratory for Data-Driven Fluid Mechanics and Engineering Applications, Southern University of Science and Technology, Shenzhen 518055, PR China
}

\begin{document}
\maketitle

\begin{abstract}
Data reconstruction of rotating turbulent snapshots is investigated utilizing data-driven tools. This problem is crucial for numerous geophysical applications and fundamental aspects, given the concurrent effects of direct and inverse energy cascades, which lead to non-Gaussian statistics at both large and small scales. Data assimilation also serves as a tool to rank physical features within turbulence, by evaluating the performance of reconstruction in terms of the quality and quantity of the information used. Additionally, benchmarking various reconstruction techniques is essential to assess the trade-off between quantitative supremacy, implementation complexity, and explicability. In this study, we use linear and non-linear tools based on the Proper Orthogonal Decomposition (POD) and Generative Adversarial Network (GAN) for reconstructing rotating turbulence snapshots with spatial damages (inpainting). We focus on accurately reproducing both statistical properties and instantaneous velocity fields. 
Different gap sizes and gap geometries are investigated in order to assess the importance of coherency and multi-scale properties of the missing information. Surprisingly enough, concerning point-wise reconstruction, the non-linear GAN does not outperform one of the linear POD techniques. On the other hand, supremacy of the GAN approach is shown when the statistical multi-scale properties are compared. Similarly, extreme events in the gap region are better predicted when using GAN. 
The balance between point-wise error and statistical properties is controlled by the adversarial ratio, which determines the relative importance of the generator and the discriminator in the GAN training. Robustness against the measurement noise is also discussed. 
\end{abstract}

\begin{keywords}
Authors should not enter keywords on the manuscript, as these must be chosen by the author during the online submission process and will then be added during the typesetting process (see \href{https://www.cambridge.org/core/journals/journal-of-fluid-mechanics/information/list-of-keywords}{Keyword PDF} for the full list).  Other classifications will be added at the same time.
\end{keywords}

{\bf MSC Codes }  {\it(Optional)} Please enter your MSC Codes here

\section{Introduction}
\label{sec:intro}
The problem of reconstructing missing information, due to measurements constraints and lack of spatial/temporal resolution, is ubiquitous in almost all important applications of turbulence to laboratory experiments, geophysics, meteorology and oceanography \citep{asch2016data, le1986variational, torn2009ensemble, bell2009godae, krysta2011consistent}. For example, satellite imagery often suffers from missing data due to dead pixels and thick cloud cover \citep{shen2015missing, zhang2018missing, militino2019filling, storer2022global}. In Particle Tracking Velocimetry (PTV) experiments \citep{dabiri2020particle}, spatial gaps naturally occur due to the use of a small number of seeded particles. Additionally, in Particle Image Velocimetry (PIV) experiments, missing information can arise due to out-of-pair particles, object shadows, or light reflection issues \citep{garcia2011fast, wang2016divergence, wen2019missing}. Similarly, in many instances, the experimental probes are limited to assess only a subset of the relevant fields, asking for a careful apriori engineering of the most relevant features to be tracked. Recently, many data-driven Machine Learning tools have been proposed to fulfil some of the previous tasks. Research using these {\it black-box} tools is at its infancy and we lack systematic quantitative benchmarks for paradigmatic high-quality and high-quantity multi-scale complex datasets, a mandatory step to make them useful for the fluid-dynamics community. In this paper, we perform a systematic quantitative comparison among three data-driven methods (no information on the underlying equations) to reconstruct highly complex two-dimensional (2D) fields from a typical geophysical set-up, as the one of rotating turbulence. The first two methods are connected with a linear 
model reduction, the so-called Proper Orthogonal Decomposition (POD) and the third is based on a fully non-linear 
Convolutional Neural Network (CNN) embedding in a framework of Generative Adversarial Network (GAN) \citep{goodfellow2014generative, deng2019super, subramaniam2020turbulence, buzzicotti2021reconstruction, kim2021unsupervised, guastoni2021convolutional, yousif2022deep, buzzicotti2022inferring}. 	POD is widely used for pattern recognition \citep{sirovich1987low, fukunaga2013introduction}, optimization \citep{singh2001optimal} and data assimilation \citep{romain2014bayesian, suzuki2014pod}. To repair the missing data in a gappy field, \citet{everson1995karhunen} proposed GPOD, where coefficients are optimized according to the measured data outside the gap. By introducing some modifications to GPOD, \citet{venturi2004gappy} improved its robustness and made it reach the maximum possible resolution at a given level of spatio-temporal gappiness. \citet{gunes2006gappy} showed that GPOD reconstruction outperfroms the Kriging interpolation \citep{oliver1990kriging, myers2002interpolation, gunes2008use}. However, GPOD is essentially a linear interpolation and thus is in trouble when dealing with complex multi-scale and non-Gaussian flows as the ones typical of fully developed turbulence \citep{alexakis2018cascades} and/or large missing areas \citep{Li2021gappy}.

EPOD was first used in \citet{maurel2001extended} on the PIV data of a turbulent internal engine flow, where the POD analysis is conducted in a sub-domain spanning only the central rotating region but the preferred directions of the jet-vortex interaction can be clearly identified. \citet{boree2003extended} generalized the EPOD and reported that EPOD can be applied to study the correlation of any physical quantity in any domain with the projection of any measured quantity on its POD modes in the measurement domain. EPOD has many applications of flow sensing, where flow predictions are made based on remote probes \citep{tinney2008low, hosseini2016modal, discetti2019characterization}. For example, using EPOD as a reference of their CNN models, \citet{guastoni2021convolutional} predicted the 2D velocity-fluctuation fields at different wall-normal locations from the wall-shear-stress components and the wall pressure in a turbulent open-channel flow. EPOD also provides a linear relation between the input and output fields.

In recent years, CNNs have made a great success in computer vision tasks \citep{niu2012novel, russakovsky2015imagenet, he2016deep} because of their powerful ability of handling nonlinearities \citep{hornik1991approximation, kreinovich1991arbitrary, baral2018deep}. In fluid mechanics, CNN has also been shown as a encouraging technique for data prediction/reconstruction \citep{fukami2019super, guemes2019sensing, kim2020prediction, https://doi.org/10.48550/arxiv.2301.07541}. Many researches devote to the super-resolution task, where CNNs are used to reconstruct high-resolution data from low-resolution data of laminar and turbulent flows \citep{liu2020deep, subramaniam2020turbulence, fukami2021machine, kim2021unsupervised}. In the scenario where a large gap exists, missing both large- and small-scale features, \citet{buzzicotti2021reconstruction} reconstructed for the first time a set of 2D damaged snapshots of three-dimensional (3D) rotating turbulence with GAN. Recent works show that CNN or GAN is also feasible to reconstruct the 3D velocity fields with 2D observations \citep{matsuo2021supervised, yousif2022deep}. GAN consists of two CNNs, a generator and a discriminator. Previous preliminary researches indicate that the introduction of discriminator 
significantly improves the high-order statistics of the prediction \citep{deng2019super, subramaniam2020turbulence, buzzicotti2021reconstruction, kim2021unsupervised, guemes2021coarse}. At difference from the previous work \citep{buzzicotti2021reconstruction}, here we attack the problem of data reconstruction with GAN at changing the ratio between the input measurements and the missing information. Furthermore, we present a first systematic assessment of the non-linear vs. linear reconstruction methods, by showing also results using two different POD-based methods. We discuss and present novel results concerning both point-based and statistical metrics. Moreover, the dependency of GAN properties on the adversarial ratio is also systematically studied. The adversarial ratio gauges the relative importance of the discriminator in comparison to the generator throughout the training process.

\begin{figure}
	\hspace{0.17\linewidth}
	\includegraphics[width=0.7\linewidth]{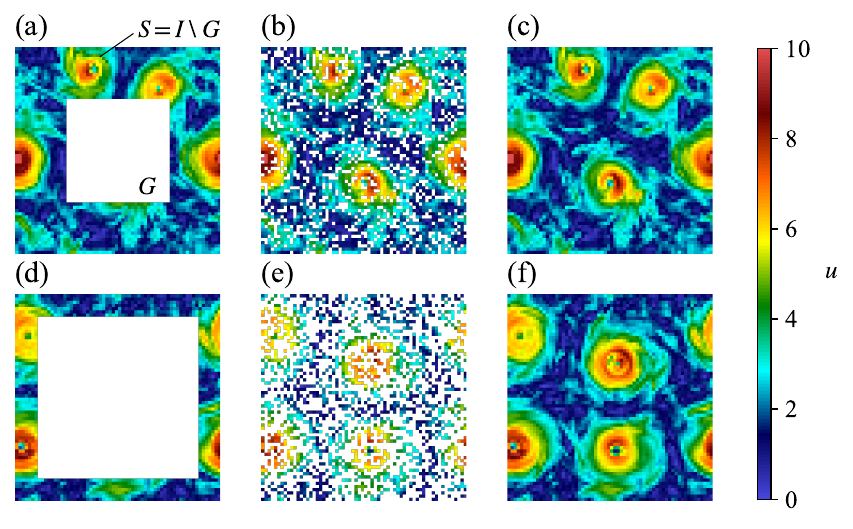}
	\caption{Examples of the reconstruction task on 2D slices of 3D turbulent rotating flows, where the color code is proportional to the velocity module. Two gap geometries are considered: (a)(d) a central square gap and (b)(e) random gappiness. Gaps in each row have the same area and the corresponding ground truth is shown in (c) and (f). We denote $G$ as the gap region and $S=I\setminus G$ as the known region, where $I$ is the entire 2D image.}
	\label{fig:Reconstruction_examples}
\end{figure}

Two factors make the reconstruction difficult. First, turbulent flows have a large number of active degrees of freedom which grows with the turbulent intensity, typically parameterized by the Reynolds number. The second factor is the spatio-temporal gappiness, which depends on the area and geometry of the missing region. In the current work we conduct a first systematic comparative study between GPOD, EPOD and GAN on the reconstruction of turbulence in the presence of rotation, which is a paradigmatic system with both coherent vortices at large scales and strong non-Gaussian and intermittent fluctuations at small scales \citep{alexakis2018cascades,buzzicotti2018inverse,di2020phase}. Figure \ref{fig:Reconstruction_examples} displays some examples of the reconstruction task in this work. The aim is to fill the gap region with data close to the ground truth (figure \ref{fig:Reconstruction_examples}(c) and (f)). A second long term goal would also be to systematically perform features ranking: understanding the quality of the supplied information on the basis of its performance in the reconstruction goal. The latter is connected to the {\it sacred grail} of turbulence: identifying the master degrees of freedoms driving turbulent flow, connected also to control problems \citep{choi1994active, lee1997application, gad2006transition, brunton2015closed, fahland2021investigation}. The study presented in this work is a first step towards a quantitative assessment of the tools that can be employed to ask and answer this kind of questions. 
In order to focus on two paradigmatic realistic set-ups, we study two gap geometries, a central square gap (figure \ref{fig:Reconstruction_examples}(a,d)) and random gappiness (figure \ref{fig:Reconstruction_examples}(b,e)). The latter is related to practical applications such as PTV and PIV. The gap area is also varied from a small to an extremely large proportion up to the limit where only one thin layer is supplied at the border, a seemingly impossible reconstructing task, for evaluation of the three methods on different situations. In a recent work, \citet{di2022reconstructing} used Physics-Informed Neural Networks (PINNs) for reconstruction with sparse and noisy particle tracks obtained experimentally. As in practice the measurements are always noisy or filtered, we also investigate the robustness of the EPOD and GAN reconstruction methods.

The paper is organized as follows. Section \ref{subsec:dataset} describes the 
dataset and the reconstruction problem set-up. The GPOD, EPOD and GAN-based reconstruction methods are introduced in \S \ref{subsec:gappy}, \S \ref{subsec:epod} and \S \ref{subsec:ganba}, respectively. In \S \ref{sec:compa}, the performances of POD- and GAN-based methods on turbulence reconstruction are systematically compared when there is one central square gap of different sizes. We address the dependency on the adversarial ratio for the GAN-based reconstruction in \S \ref{sec:depen} and show results for random gappiness from GPOD, EPOD and GAN in \S \ref{sec:rando}. The robustness of EPOD and GAN to measurement noise and the computational cost of all methods are discussed in \S\ref{sec:discu}. Finally, conclusions of the work are presented in \S \ref{sec:concl}.

\section{Methodology}
\subsection{Dataset and reconstruction problem set-up}
\label{subsec:dataset}
\begin{figure}
	\centering
	\includegraphics[width=1.0\linewidth]{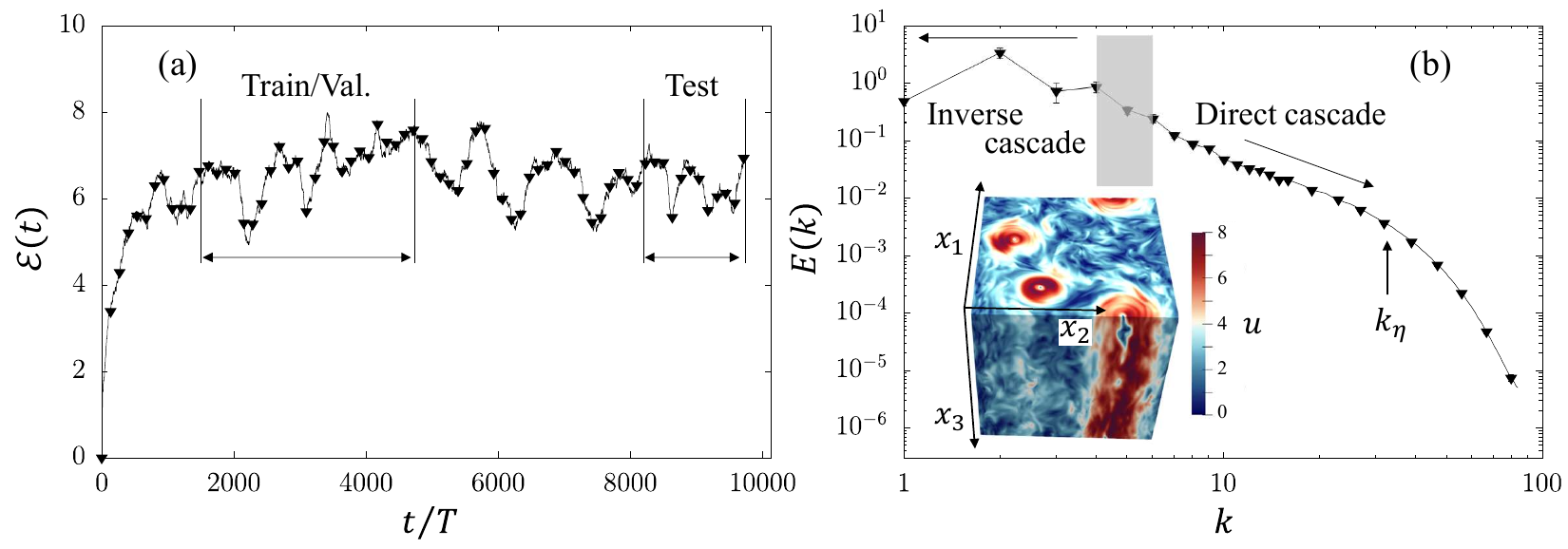}
	\caption{(a) Energy evolution of the turbulent flow, where we also show the sampling time periods for the training/validation and testing data. (b) The averaged energy spectrum. The gray area indicates the forcing wave numbers and $k_\eta$ is the Kolmogorov dissipative wave number where $E(k)$ starts to decay exponentially. The inset shows an instantaneous visualization of velocity module with the frame of reference for the simulation.}
	\label{fig:Dataset}
\end{figure}

For the evaluation of different reconstruction tools, we use a dataset from the TURB-Rot \citep{biferale2020turb} open database. The dataset used in this study is generated from a direct numerical simulation (DNS) of the Navier-Stokes equations for the homogeneous incompressible flow in the presence of rotation with periodic boundary conditions \citep{godeferd2015structure, seshasayanan2018condensates, pouquet2018scaling, van2020critical, yokoyama2021energy}. 
In a rotating frame of reference, both Coriolis and centripetal accelerations must be taken into account. However, the centrifugal force can be expressed as the gradient of the centrifugal potential and included in the pressure gradient term. In this way, the resulting equations will explicitly show only the presence of the Coriolis force, while the centripetal term is absorbed into a modified pressure \citep{cohen2004fluid}. 
The simulation is performed using a fully dealiased parallel pseudo-spectral code in a 3D ($x_1$-$x_2$-$x_3$) periodic domain of size $[0,2\pi)^3$ with 256 grid points in each direction, as shown in the inset of figure \ref{fig:Dataset}(b). Denote $l_0=2\pi$ as the domain size, the Fourier spectral wave number is $\bm{k}=(k_1,k_2,k_3)$, where $k_1=2n_1\pi/l_0$ ($n_1\in\mathbb{Z}$) and one can similarly obtain $k_2$ and $k_3$. The governing equations can be written as
\begin{equation}\label{equ:mmt}
	\frac{\p\bm{u}}{\p t}+\bm{u}\cdot\bnabla\bm{u}+2\bm{\mathit{\Omega}}\times\bm{u}=-\frac{1}{\rho}\bnabla\tilde{p}+\nu\nabla^2\bm{u}+\bm{f},
\end{equation}
where $\bm{u}$ is the incompressible velocity, $\bm{\mathit{\Omega}}=\mathit{\Omega}\hat{\bm{x}}_3$ is the system rotation vector, $\tilde{p}=p+\frac{1}{2}\rho\lVert\bm{\Omega}\times\bm{x}\rVert^2$ is the pressure in an inertial frame modified by a centrifugal term, $\nu$ is the kinematic viscosity and $\bm{f}$ is an external forcing mechanism at scales around $k_f=4$ via a second-order Ornstein-Uhlenbeck process \citep{sawford1991reynolds, buzzicotti2016lagrangian}. Figure \ref{fig:Dataset}(a) plots the energy evolution with time of the whole simulation. The energy spectrum $E(k)=\frac{1}{2}\sum_{k\le\lVert\bm{k}\rVert<k+1}\lVert\hat{\bm{u}}(\bm{k})\rVert^2$ averaged over time is shown in figure \ref{fig:Dataset}(b), where the gray area indicates the forcing wave numbers. To enlarge the inertial range between $k_f$ and the Kolmogorov dissipative wave number, $k_\eta=32$, which is picked as the scale where $E(k)$ starts to decay exponentially, the viscous term $\nu\nabla^2\bm{u}$ in equation (\ref{equ:mmt}) is replaced with a hyperviscous term $\nu_h\nabla^4\bm{u}$ \citep{haugen2004inertial, frisch2008hyperviscosity}. 
We define an effective Reynolds number as $Re_\mathrm{eff}=(k_0/k_\eta)^{-3/4}\approx13.45$, with the smallest wave number $k_0=1$. A linear friction term $-\beta\bm{u}$ acting only on wave numbers $\lVert\bm{k}\rVert\le2$ is also used in r.h.s. of (\ref{equ:mmt}) to prevent a large-scale condensation \citep{alexakis2018cascades}. As shown in figure \ref{fig:Dataset}(a), the flow reaches a stationary state with a Rossby number $Ro=\mathcal{E}^{1/2}/(\mathit{\Omega}/k_f)\approx0.1$, where $\mathcal{E}$ is the kinetic energy. The integral length scale is $L=\mathcal{E}/\int kE(k)\,\mathrm{d}k\sim0.15l_0$ and the integral time scale is $T=L/\mathcal{E}^{1/2}\approx0.185$. Readers can refer to \citet{biferale2020turb} for more details on the simulation.

The dataset is extracted from the above simulation as follows:

First, we sampled 600 snapshots of the whole 3D velocity field from time $t=276$ up to $t=876$ for training and validation, and we sampled 160 3D snapshots from $t=1516$ to $t=1676$ for testing, as shown in figure \ref{fig:Dataset}(a). A sampling interval $\Delta t_s\approx5.41T$ is used to decrease correlations in time between two successive snapshots.

 To reduce the amount of data to be analyzed, the resolution of sampled fields is downsized from $256^3$ to $64^3$ by a spectral low-pass filter
\begin{equation}\label{equ:filter}
    \bm{u}(\bm{x})=\sum_{\lVert\bm{k}\rVert\le k_\eta}\hat{\bm{u}}(\bm{k})\mathrm{e}^{\mathrm{i}\bm{k}\cdot\bm{x}},
\end{equation}
where the cut-off is the Kolmogorov dissipative wave number such as to only eliminate the fully dissipative degrees of freedom where the flow becomes smooth \citep{frisch1995turbulence}. Therefore, there is not a loss of data complexity in this procedure and it also indicates that the measurement resolution corresponds to $k_\eta$.

For each downsized 3D field, we selected 16 horizontal ($x_1$-$x_2$) planes at different $x_3$, each of which can be augmented to 11 (for training and validation) or 8 (for testing) different ones by randomly shifting it along both $x_1$ and $x_2$ directions using periodic boundary conditions \citep{biferale2020turb}. Therefore, a total of 176 or 128 planes can be obtained at each instant of time.

Finally, the 105600 planes sampled at early times are randomly shuffled and used to constitute the Train/Validation split: 84480 (80\%)/10560 (10\%), which is used for the training process, while the other 20480 planes sampled at later times are used for the testing process.

The parameters of the dataset are summarized in table \ref{tab:Dataset}. In the present study, we only reconstruct the velocity module, $u=\lVert\bm{u}\rVert$, which is always positive. Note that we restrict our study to 2D horizontal slices in order to make contact with geophysical observation, although GPOD, EPOD and GAN are feasible to 3D data.
\begin{table}
	\begin{center}
		\def~{\hphantom{0}}
		\begin{tabular}{cccccccccc}
			$Re_\mathrm{eff}$ & $Ro$ & $L$ & $N_{x_1}\times N_{x_2}$ & $\Delta t_s/T$ & $N_\mathrm{train}$ & $N_\mathrm{valid}$ & $T_\mathrm{train}/T$ & $N_\mathrm{test}$ & $T_\mathrm{test}/T$ \\[3pt]
			13.45 & 0.1 & $0.15l_0$ & $64\times64$ & 5.41 & 84480 & 10560 & 3243 & 20480 & 865\\
		\end{tabular}
		\caption{Description of the dataset used for the evaluation of reconstruction methods. Here, $N_{x_1}$ and $N_{x_2}$ indicate the resolution of the horizontal plane. The number of fields for training/validation/testing is denoted as $N_\mathrm{train}$/$N_\mathrm{valid}$/$N_\mathrm{test}$. The sampling time periods for training/validation and testing are $T_\mathrm{train}$ and $T_\mathrm{test}$, respectively.}
		\label{tab:Dataset}
	\end{center}
\end{table}

\begin{figure}
	\centering
	\begin{tikzpicture}[scale=0.5]
		\draw[thick] (0,0) rectangle (7,7);
		\draw[fill=gray!50] plot[smooth cycle] coordinates {(3,3.5) (4,4.5) (5,3.5) (5.5,4) (6,3.5) (5.5,2.5) (6.5,1.5) (6.5,1) (5.5,1) (5.5,2) (3.5,2.5)};
		\draw[fill=gray!50] plot[smooth cycle] coordinates {(.5,.5) (1,2) (2.5,3) (2,1)};

		\node at (6.5,7) [above] {\(I\)};
		\draw (2,2) -- (3.5,1.5) node[fill=white] {\(G\)} -- (4,3);
		\node at (0,7) [below right] {\(S = I \setminus G\)};

		\node at (3,6) [below right] {\large \(u_S(\bm{x})\)};
        \draw (5.5,3.5) -- (8.8,3.5) node[fill=white] {\large \(u_G^{(t)}(\bm{x})\)};
	\end{tikzpicture}
	\caption{Schematic diagram of a gappy field.}
	\label{fig:Gappy_field}
\end{figure}
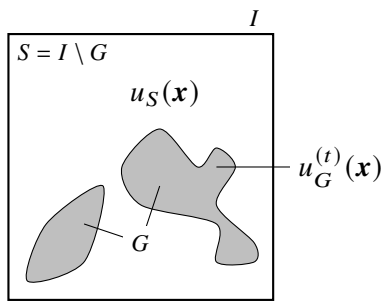

We next describe the reconstruction problem set-up. Figure \ref{fig:Gappy_field} presents an example of a gappy field, where $I$, $G$ and $S$ represent the whole region, the gap region and the known region, respectively. Given the damaged area $A_G$, we can define the gap size as $l=\sqrt{A_G}$. As shown in figure \ref{fig:Reconstruction_examples}, two gap geometries are considered: i) a square gap located at the center and ii) random gappiness which spreads over the whole region. Once the positions in $G$ are determined, $G$ is fixed for all planes over the training and the testing processes. 
Note that the GAN-based reconstruction can also handle the case where $G$ is randomly changed for different planes (not shown). For a field $u(\bm{x})$ defined on $I$, we define the supplied measurements in $S$ as $u_S(\bm{x})=u(\bm{x})$ (with $\bm{x}\in S$), and the ground truth or the predicted field in $G$, as $u_G^{(t)}(\bm{x})=u(\bm{x})$ or $u_G^{(p)}(\bm{x})$ (with $\bm{x}\in G$). 
The reconstruction models are `learned' with the training data defined on the whole region $I$. Once the training process completed, one can evaluate the models by comparing the prediction and the ground truth in $G$ over the test dataset.


\subsection{GPOD reconstruction}
\label{subsec:gappy}
This section briefly presents the procedure of GPOD. The first step is to conduct POD analysis with the training data on the whole region $I$, namely solving the eigenvalue problem
\begin{equation}\label{equ:eig_prob1}
	\int_I R_I(\bm{x},\bm{y})\psi_n(\bm{y})\,\mathrm{d}\bm{y}=\lambda_n\psi_n(\bm{x})\qquad(\bm{x}\in I),
\end{equation}
where
\begin{equation}
	R_I(\bm{x},\bm{y})=\langle u(\bm{x})u(\bm{y})\rangle\qquad(\bm{x},\bm{y}\in I)
\end{equation}
is the correlation matrix, given $\langle\cdot\rangle$ as the average over training dataset. We denote $\lambda_n$ as the eigenvalues and $\psi_n(\bm{x})$ as the POD eigenmodes, where $n=1,\ldots,N_I$ and $N_I=N_{x_1}\times N_{x_2}$ is the number of points in $I$. 
For the homogeneous periodic flow considered in this study, it can be demonstrated that the POD modes correspond to Fourier modes, and their spectra are identical \citep{holmes2012turbulence}.
In all POD analyses of the present study, the mean of $u(\bm{x})$ is not removed. Any realization of the field can be decomposed as
\begin{equation}\label{equ:poddec}
	u(\bm{x})=\sum_{n=1}^{N_I}a_n\psi_n(\bm{x})\qquad(\bm{x}\in I)
\end{equation}
with the POD coefficients
\begin{equation}
\label{eq:apod}
	a_n=\int_I u(\bm{x})\psi_n(\bm{x})\,\mathrm{d}\bm{x}.
\end{equation}
In the case when we have data only in $S$, the relation (\ref{eq:apod}) cannot be used and one can adopt the dimension reduction by keeping only the first $N'$ POD modes and minimize the distance between the measurements and the linear POD decomposition \citep{everson1995karhunen},
\begin{equation}\label{equ:err_gpod}
	\tilde{E} = \int_S | u_S(\bm{x}) - \sum_{n=1}^{N'} a_{n}^{(p)}\psi_n(\bm{x}) |^2 \, \mathrm{d}\bm{x},
\end{equation}
to obtain the predicted coefficients $\{a_{n}^{(p)}\}_{n=1}^{N'}$. Then the GPOD prediction can be given as
\begin{equation}\label{equ:gpod}
	u_G^{(p)}(\bm{x})=\sum_{n=1}^{N'}a_{n}^{(p)}\psi_n(\bm{x})\qquad(\bm{x}\in G).
\end{equation}
We optimize the value of $N'$ during the training phase by requiring a minimum  mean $L_2$ distance with the ground truth in the gap: 
\begin{equation}\label{equ:Np}
	\argmin_{N'}\langle\int_G |u_G^{(t)}(\bm{x})-u_G^{(p)}(\bm{x})|^2\,\mathrm{d}\bm{x}\rangle.
\end{equation}
Table \ref{tab:Np} summarizes the optimal $N'$ used in this study.
\begin{table}
	\begin{center}
		\def~{\hphantom{0}}
			\begin{tabular}{ccccccccc}
				$l$ & 8 & 16 & 24 & 32 & 40 & 50 & 60 & 62 \\[3pt]
				$N'$ (s. g.) & 72 & 45 & 21 & 21 & 13 & 13 & 13 & 13\\
				$N'$ (r. g.) & 2334 & 2069 & 1726 & 1403 & 1039 & 551 & 98 & 56\\
			\end{tabular}
		\caption{Summary of the optimal $N'$ for the square gap (s. g.) and random gappiness (r. g.) with different sizes.}
		\label{tab:Np}
	\end{center}
\end{table}
An analysis of reconstruction error for different $N'$ is conducted in Appendix \ref{appA}. Let us notice that there also exists a different approach to select, frame-by-frame, a subset of POD modes to be used in the GPOD approach, based on Lasso, a regression analysis that performs mode selection with regularization \citep{tibshirani1996regression}. Results using this second approach do not show any significant improvement in a typical case of our study (see Appendix \ref{appB}).

\subsection{EPOD reconstruction}\label{subsec:epod}
To use EPOD for flow reconstruction, we first compute the correlation matrix
\begin{equation}
	R_S(\bm{x},\bm{y})=\langle u_S(\bm{x})u_S(\bm{y})\rangle\qquad(\bm{x},\bm{y}\in S)
\end{equation}
and solve the eigenvalue problem
\begin{equation}\label{equ:eig_prob2}
	\int_S R_S(\bm{x},\bm{y})\phi_n(\bm{y})\,\mathrm{d}\bm{y}=\sigma_n\phi_n(\bm{x})\qquad(\bm{x}\in S)
\end{equation}
to obtain the eigenvalues $\sigma_n$ and the POD eigenmodes $\phi_n(\bm{x})$, where $n=1,\ldots,N_S$ and $N_S$ equals to the number of points in $S$. 
We remark that $\phi_n(\bm{x})$ are not Fourier modes, as the presence of the internal gap breaks the homogeneity. 
Any realization of the measured field in $S$ can be decomposed as
\begin{equation}\label{equ:dec}
	u_S(\bm{x})=\sum_{n=1}^{N_S}b_n\phi_n(\bm{x})\qquad(\bm{x}\in S),
\end{equation}
where the $n$-th POD coefficient is obtained from
\begin{equation}\label{equ:coe}
	b_n=\int_S u_S(\bm{x})\phi_n(\bm{x})\,\mathrm{d}\bm{x}.
\end{equation}
Furthermore, with (\ref{equ:dec}) and an important property \citep{boree2003extended}, $\langle b_nb_p\rangle=\sigma_n\delta_{np}$, one can derive the following identity:
\begin{equation}\label{equ:POD}
	\phi_n(\bm{x})=\frac{\langle b_n u_S(\bm{x})\rangle}{\sigma_n}\qquad(\bm{x}\in S).
\end{equation}
Here, we reiterate that $\langle\cdot\rangle$ denotes the average over the training dataset. Specifically, $\langle b_n u_S\rangle$ can be interpreted as $(1/N_\mathrm{train})\sum_{c=1}^{N_\mathrm{train}}b_n^{(c)}u_S^{(c)}$, where the superscript $c$ represents the index of a particular snapshot.
The {\it Extended} POD mode is defined by replacing $u_S(\bm{x})$ with the field to be predicted $u_G^{(t)}(\bm{x})$ in (\ref{equ:POD}):
\begin{equation}\label{equ:ext}
	\phi_n^E(\bm{x})=\frac{\langle b_n u_G^{(t)}(\bm{x})\rangle}{\sigma_n}\qquad(\bm{x}\in G).
\end{equation}
Once obtained the set of EPOD modes (\ref{equ:ext}) in the training process one can start the reconstruction of a test data with the measurement $u_S(\bm{x})$ from calculating the POD coefficients (\ref{equ:coe}) and the prediction in $G$ can be obtained as the correlated part with $u_S(\bm{x})$ \citep{boree2003extended}:
\begin{equation}\label{equ:epod}
	u_G^{(p)}(\bm{x})=\sum_{n=1}^{N_S}b_n\phi_n^E(\bm{x})\qquad(\bm{x}\in G).
\end{equation}

\subsection{GAN-based reconstruction with Context Encoders}
\label{subsec:ganba}
\begin{figure}
	\centering
	\includegraphics[width=0.95\linewidth]{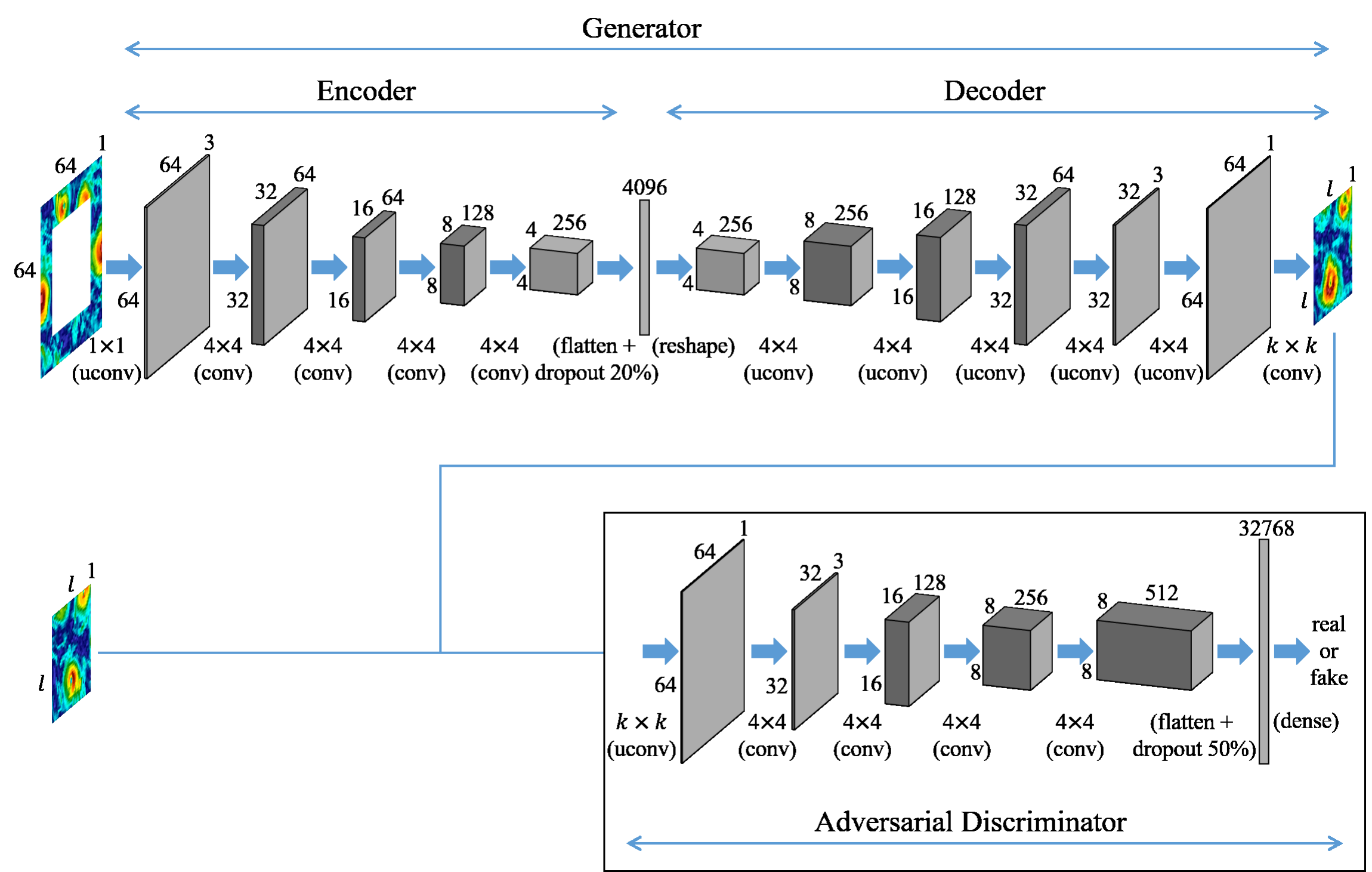}
	\caption{Architecture of generator and discriminator network for flow reconstruction with a square gap. The kernel size $k$ and the corresponding stride $s$ are determined based on the gap size $l$. Similar architecture holds for random gappiness as well.}
	\label{fig:GAN_schematic_diagram}
\end{figure}

In a previous work, \citet{buzzicotti2021reconstruction} used a context encoder \citep{pathak2016context} embedding in GAN to generate missing data for the case where the total gap size is fixed, but with different spatial distributions. To
generalize the previous approach to study gaps of different geometries and sizes, here we 
extend previous GAN architecture by adding one layer at the start, 
two layers at the end of the generator and one layer at the start of the discriminator, as shown in figure \ref{fig:GAN_schematic_diagram}. The generator is a functional $GEN(\cdot)$ first taking the damaged `context', $u_S(\bm{x})$, to produce a latent feature representation with an encoder, and second with a decoder 
to predict the missing data, $u_G^{(p)}(\bm{x})=GEN(u_S(\bm{x}))$. The latent feature represents the output vector of the encoder with $4096$ neurons in figure \ref{fig:GAN_schematic_diagram}, 
extracted from the input with the convolutions and nonlinear activations. To constrain the predicted velocity module being positive, a Rectified Linear Unit (ReLU) activation function is adopted at the last layer of generator. The discriminator acts as a `referee' functional $D(\cdot)$, which takes either $u_G^{(t)}(\bm{x})$ or $u_G^{(p)}(\bm{x})$ and outputs the probability that the provided input ($u_G^{(t)}$ or $u_G^{(p)}$) belongs to the real turbulent ensemble.
The generator is trained to minimize the following loss function:
\begin{equation}\label{equ:L_GAN}
	\mathcal{L}_{GEN}=(1-\lambda_{adv})\mathcal{L}_\mathrm{MSE}+\lambda_{adv}\mathcal{L}_{adv},
\end{equation}
where the $L_2$ loss
\begin{equation}
	\label{eq:GAN}
	\mathcal{L}_\mathrm{MSE}=\langle\frac{1}{A_G}\int_G|u_G^{(p)}(\bm{x})-u_G^{(t)}(\bm{x})|^2\,\mathrm{d}\bm{x}\rangle
\end{equation}
is the mean squared error (MSE) between the prediction and the ground truth. 
It is important to stress that on the contrary of the GPOD case, here the supervised $L_2$ loss is calculated only inside the gap region $G$. The hyper-parameter $\lambda_{adv}$ is called the adversarial ratio and the adversarial loss is
\begin{align}\label{equ:L_adv}
	\mathcal{L}_{adv}&=\langle\log(1-D(u_G^{(p)}))\rangle=\int p(u_S)\log[1-D(GEN(u_S))]\,\mathrm{d}u_S \nonumber \\
	&=\int p_p(u_G)\log(1-D(u_G))\,\mathrm{d}u_G,
\end{align}
where $p(u_S)$ is the probability distribution of the field in $S$ over the training dataset and $p_p(u_G)$ is the probability distribution of the predicted field in $G$ given by the generator.
At the same time, the discriminator is trained to maximize the cross entropy based on its classification prediction for both real and predicted samples,
\begin{align}\label{equ:L_DIS}
	\mathcal{L}_{DIS}&=\langle\log(D(u_G^{(t)}))\rangle+\langle\log(1-D(u_G^{(p)}))\rangle \nonumber \\
	&=\int[p_t(u_G)\log(D(u_G))+p_p(u_G)\log(1-D(u_G))]\,\mathrm{d}u_G,
\end{align}
where $p_t(u_G)$ is the probability distribution of the ground truth, $u_G^{(t)}(\bm{x})$. \citet{goodfellow2014generative} further showed that the adversarial training between generator and discriminator with $\lambda_{adv}=1$ in (\ref{equ:L_GAN}) minimizes the Jensen–Shannon (JS) divergence between the real and the predicted distributions, $\mathrm{JSD}(p_t\parallel p_p)$. Refer to (\ref{equ:JSD}) for the definition of the JS divergence. Therefore, the adversarial loss helps the generator to produce predictions that are statistically similar to real turbulent configurations. It is important to stress that the adversarial ratio $\lambda_{adv}$, which controls the weighted summation of $\mathcal{L}_\mathrm{MSE}$ and $\mathcal{L}_{adv}$, is tuned to reach a balance between the MSE and turbulent statistics of the reconstruction (see \S\ref{sec:depen}). 
More details about the GAN are discussed in Appendix \ref{appC}, including the architecture, hyper-parameters and the training schedule.

\section{Comparison between POD- and GAN-based reconstructions}
\label{sec:compa}
To conduct a systematic comparison between POD- and GAN-based reconstructions, we start by studying the case with a central square gap of various sizes (see figure \ref{fig:Reconstruction_examples}). All reconstruction methods are first evaluated with the predicted velocity module itself, which is dominated by the large-scale coherent structures. The predictions are further assessed from a multi-scale perspective, with the help of the gradient of the predicted velocity module, spectral properties and multi-scale flatness. Finally, the performance on predicting extreme events is studied for all methods.

\subsection{Large-scale information}
\label{subsec:large}
\begin{figure}
	\centering
	\includegraphics[width=1.0\linewidth]{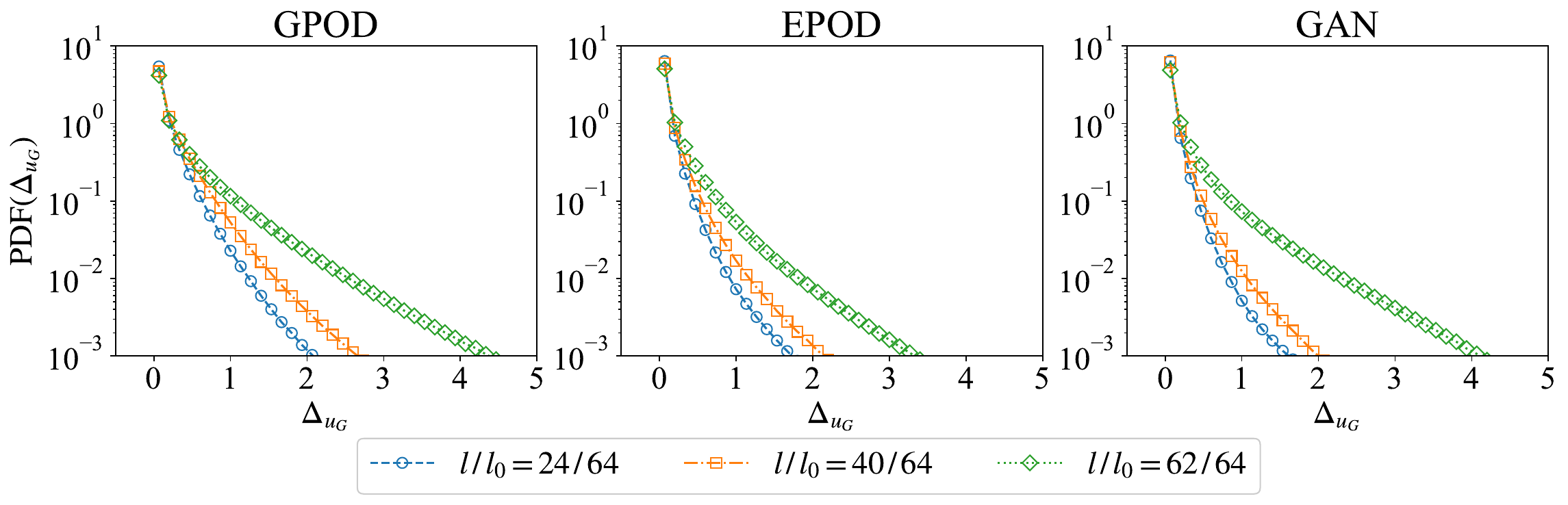}
	\includegraphics[width=1.0\linewidth]{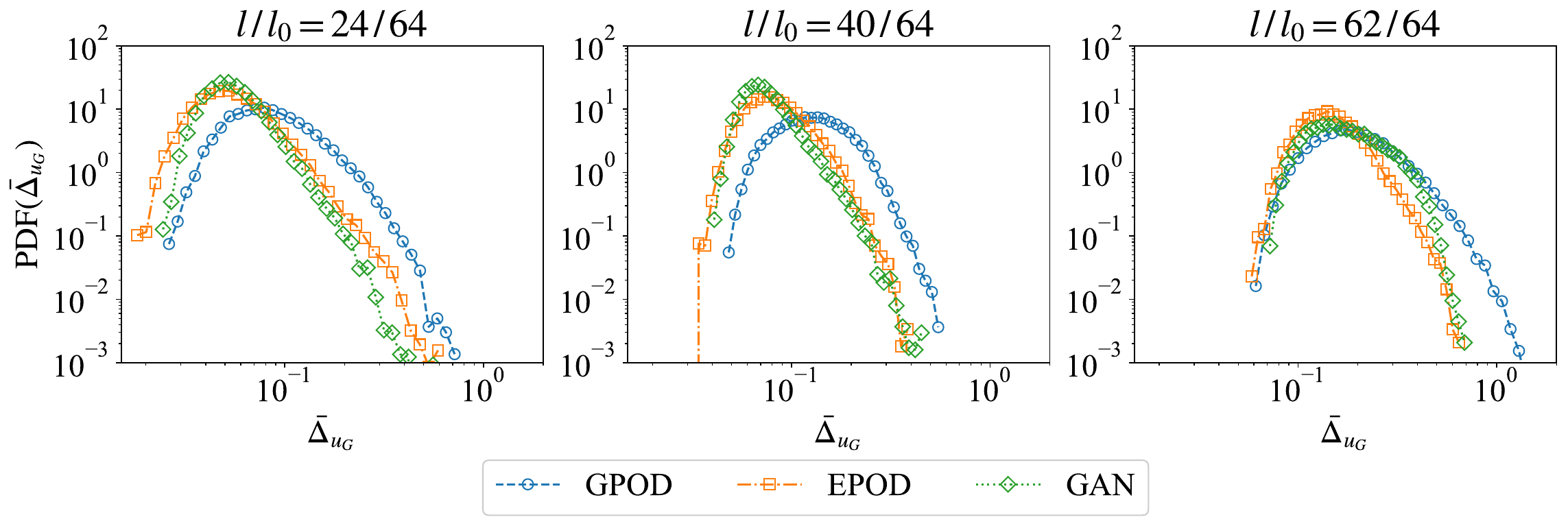}
	\caption{PDFs of the spatially averaged $L_2$ error over different flow configurations obtained from GPOD, EPOD and GAN for a square gap with sizes $l/l_0=24/64$, $40/64$ and $62/64$.}
	\label{fig:PDF_L2_c-sq_gap}
\end{figure}

\begin{figure}
	\centering
	\includegraphics[width=1.0\linewidth]{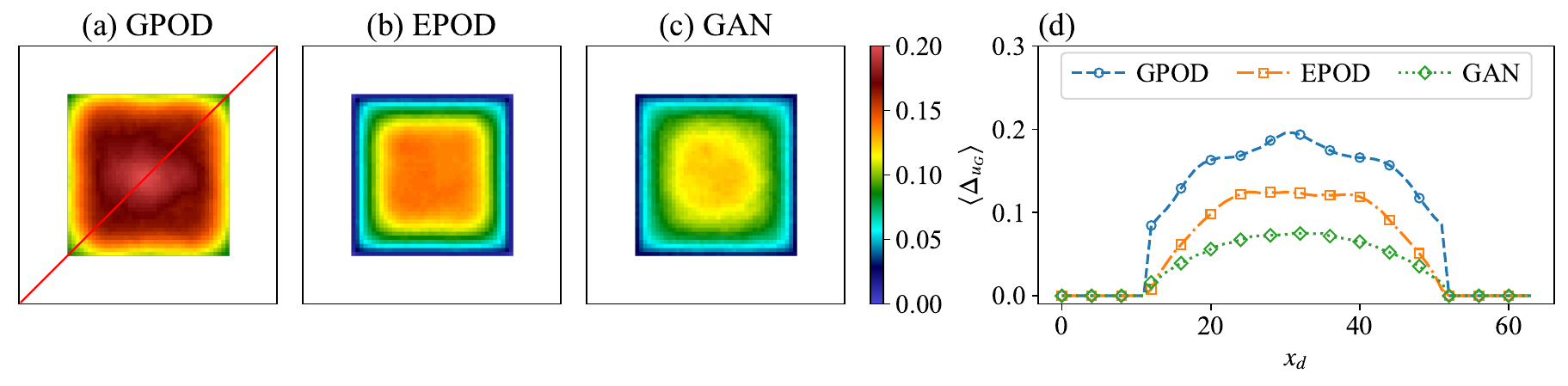}
	\caption{Averaged point-wise $L_2$ error obtained from (a) GPOD, (b) EPOD and (c) GAN for a square gap of size $l/l_0=40/64$. (d) Profiles of $\langle\Delta_{u_G}\rangle$ along the red diagonal line shown in (a), parameterized by $x_d$.}
	\label{fig:MSE_heatmap-sq_gap}
\end{figure}

\begin{figure}
	\hspace{0.07\linewidth}
	\includegraphics[width=0.9\linewidth]{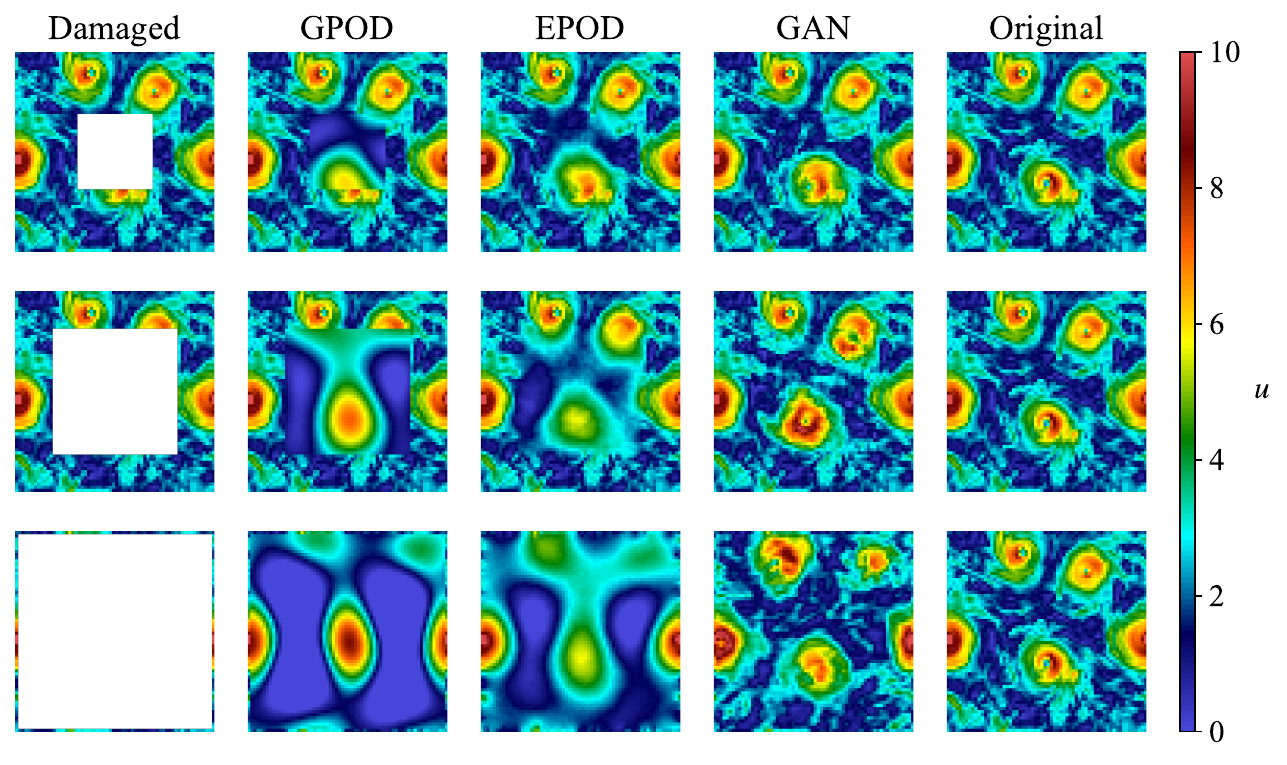}
	\caption{Reconstruction of an instantaneous field (velocity module) by the different tools for a square gap of sizes $l/l_0=24/64$ (1st row), $40/64$ (2nd row) and $62/64$ (3rd row). The damaged fields are shown in the 1st column, while the 2th to 4th columns show the reconstructed fields obtained from GPOD, EPOD and GAN. The ground truth is shown in the 5th column.}
	\label{fig:Reconstruction-sq_gap}
\end{figure}

In this section, the predicted velocity module in the missing region is quantitatively evaluated. First we consider the reconstruction error and define the normalized MSE in the gap as
\begin{equation}\label{equ:MSE}
	\mathrm{MSE}(u_G) = \frac{1}{E_{u_G}} \langle \frac{1}{A_G} \int_G | u_G^{(t)}(\bm{x}) - u_G^{(p)}(\bm{x}) |^2 \, \mathrm{d}\bm{x} \rangle,
\end{equation}
where $\langle\cdot\rangle$ represents hereafter the average over the test data. The normalization factor is defined as
\begin{equation}\label{equ:EuG}
	E_{u_G} = \sigma_G^{(p)} \sigma_G^{(t)},
\end{equation}
where
\begin{equation}
	\sigma_G^{(p)} = \langle \frac{1}{A_G} \int_G |u_G^{(p)}|^2(\bm{x}) \, \mathrm{d}\bm{x} \rangle^{1/2}
\end{equation}
and $\sigma_G^{(t)}$ is defined similarly. With the specific form of $E_{u_G}$, predictions with too small or too large energy will give a large MSE.
To provide a baseline for MSE, a set of predictions can be made by randomly sampling the missing field from the true turbulent data. In other words, the baseline comes from uncorrelated predictions that are statistically consistent with the ground truth. The baseline value is around 0.54, see Appendix \ref{appD}. Figure \ref{fig:MSE-gap_sizes-sq_gap} shows the $\mathrm{MSE}(u_G)$ from GPOD, EPOD and GAN reconstructions in a square gap with different sizes. The MSE is first calculated over data batches of size 128 (batch size used for GAN training), then the same calculation is repeated over 160 different batches, from which we calculate the MSE mean and its range of variation. 
EPOD and GAN reconstructions provide similar MSEs except at the largest gap size, where GAN has a little bit larger MSE than EPOD. Besides, both EPOD and GAN have smaller MSEs than GPOD for all gap sizes.
\begin{figure}
	\centering
	\includegraphics[width=0.7\linewidth]{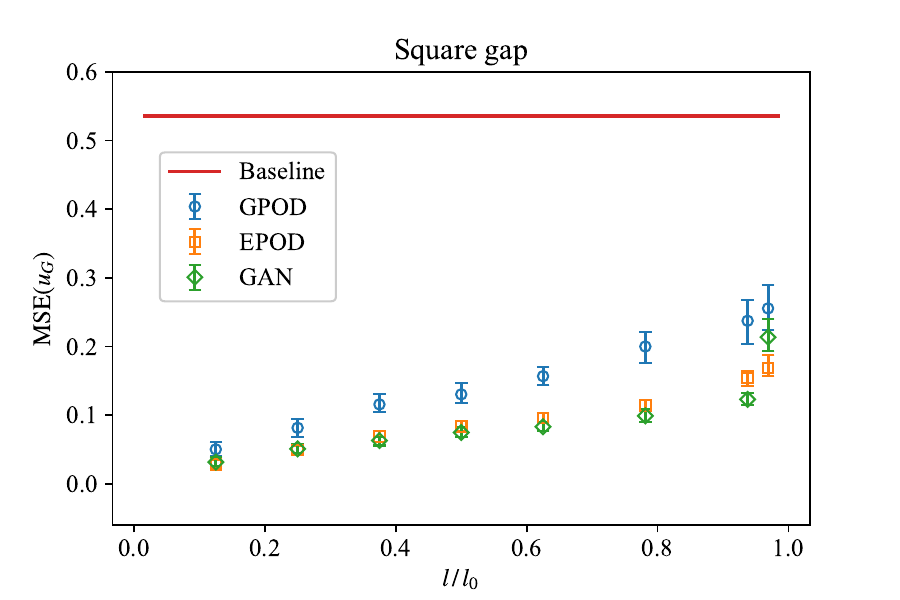}
	\caption{The MSE of the reconstructed velocity module from GPOD, EPOD and GAN in a square gap with different sizes. The abscissa is normalized by the domain size $l_0$. Red horizontal line represents the uncorrelated baseline.}
	\label{fig:MSE-gap_sizes-sq_gap}
\end{figure}
Figure \ref{fig:PDF_L2_c-sq_gap} shows the probability density function (PDF) of the spatially averaged $L_2$ error in the missing region for one flow configuration
\begin{equation}
	\bar{\Delta}_{u_G}=\frac{1}{A_G}\int_G\Delta_{u_G}(\bm{x})\,\mathrm{d}\bm{x},
\end{equation}
where
\begin{equation}
	\Delta_{u_G}(\bm{x}) = \frac{1}{E_{u_G}} | u_G^{(p)}(\bm{x}) - u_G^{(t)}(\bm{x}) |^2
\end{equation}
is the normalized point-wise $L_2$ error. The PDFs are shown for three different gap sizes $l/l_0=24/64$, $40/64$ and $62/64$. Clearly, the PDFs concentrating on regions of smaller $\bar{\Delta}_{u_G}$ correspond to the cases with smaller MSEs in figure \ref{fig:MSE-gap_sizes-sq_gap}.
To further study the performance of the three tools, we plot the averaged point-wise $L_2$ error, 
$\langle\Delta_{u_G}(\bm{x})\rangle$, for a square gap of size $l/l_0=40/64$ in figure \ref{fig:MSE_heatmap-sq_gap}. It shows that GPOD produces large $\langle\Delta_{u_G}\rangle$ all over the gap, while EPOD and GAN behave quite better, especially for the edge region. Moreover, GAN generates smaller $\langle\Delta_{u_G}\rangle$ than EPOD in the inner area (figure \ref{fig:MSE_heatmap-sq_gap}(b) and (c)).
However, the $L_2$ error is naturally dominated by the more energetic structures (the ones found at large scales in our turbulent flows) and does not provide an informative evaluation of the predicted fields at multiple scales, which is also important for assessing the reconstruction tools for the turbulent data.
Indeed, from figure \ref{fig:Reconstruction-sq_gap} it is possible to see in a glimpse that the POD- and GAN-based reconstructions have completely different multi-scale statistics which is not captured by the MSE. Figure \ref{fig:Reconstruction-sq_gap} shows predictions of an instantaneous velocity module field based on GPOD, EPOD and GAN methods compared with the ground truth solution. For all three gap sizes $l/l_0=24/64$, $40/64$ and $62/64$, GAN produces realistic reconstructions while GPOD and EPOD only generates blurry predictions. Besides, there are also obvious discontinuities between the supplied measurements and the GPOD predictions of the missing part.
This is clearly due to the fact that the number of POD modes $N'$ used for prediction in (\ref{equ:gpod}) is limited, as there are only $N_S$ measured points available in (\ref{equ:err_gpod}) for each damaged data (thus $N'<N_S$). Moreover, minimizing the $L_2$ distance from ground truth in (\ref{equ:Np}) results in solutions with almost the correct energy contents but without the complex multi-scale properties.
Unlike GPOD using global basis defined on the whole region $I$, EPOD gives better results by considering the correlation between fields defined on two smaller regions, $S$ and $G$. In this way the prediction (\ref{equ:epod}) has the degrees of freedom equal to $N_S$, which are larger than those for GPOD. Therefore, EPOD can predict the large-scale coherent structures but is still limited in generating correct multi-scale properties. Specifically, when the gap size is extremely large, $N_S$ is very small thus both GPOD and EPOD have small degrees of freedom to make realistic predictions.

\begin{figure}
	\centering
	\includegraphics[width=0.7\linewidth]{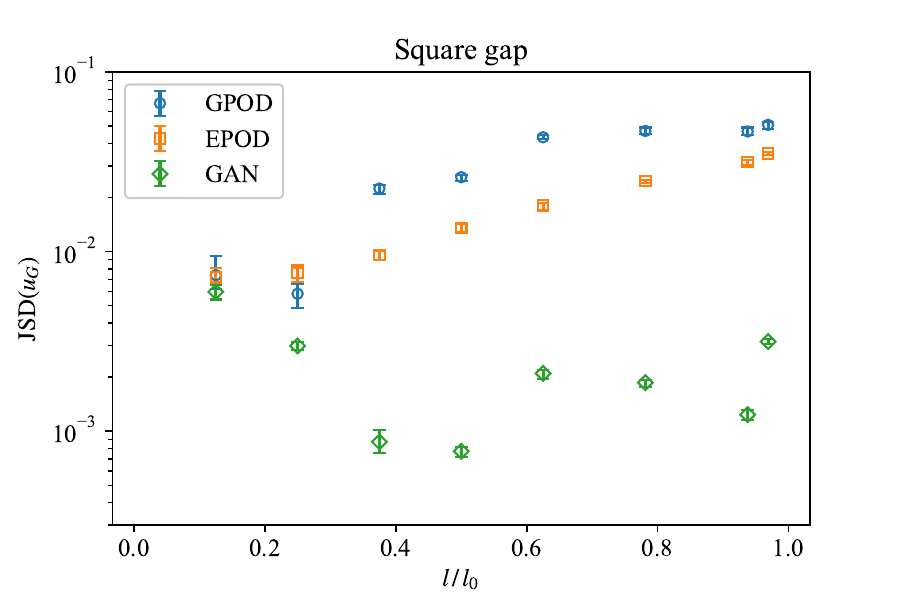}
	\caption{The JS divergence between PDFs of the velocity module inside the missing region from the original data and the predictions obtained from GPOD, EPOD and GAN for a square gap with different sizes.}
	\label{fig:JSD-gap_sizes-sq_gap}
\end{figure}

\begin{figure}
	\centering
	\includegraphics[width=1.0\linewidth]{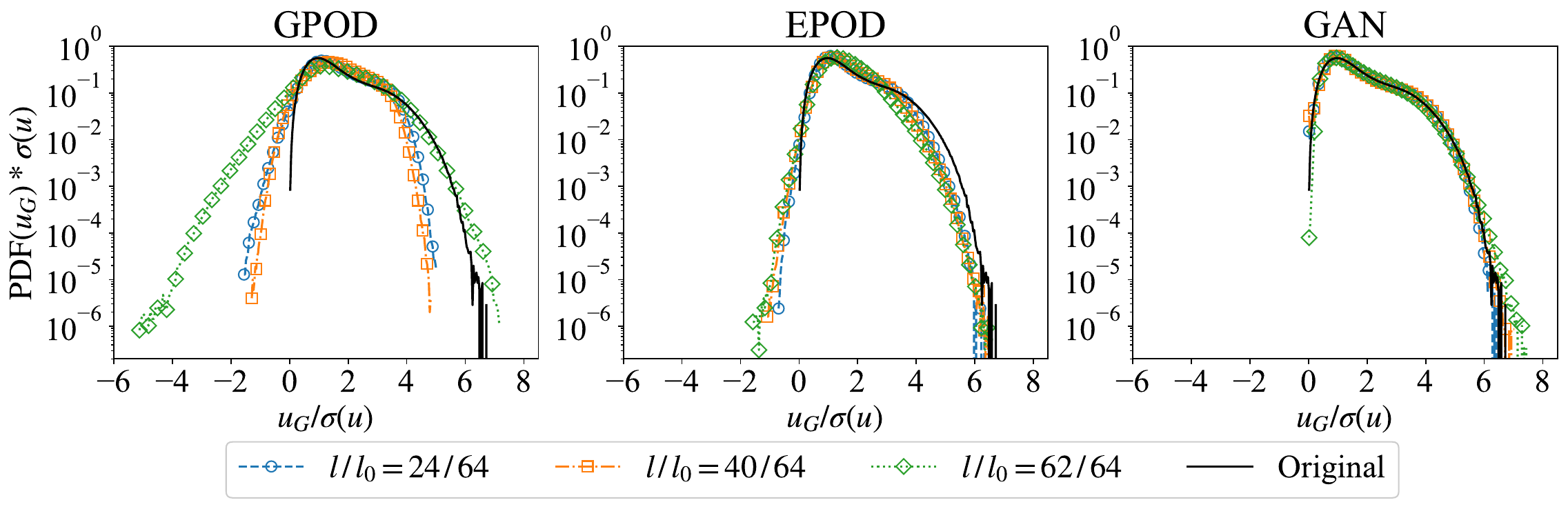}
	\caption{PDFs of the velocity module in the missing region obtained from GPOD, EPOD and GAN for a square gap with different sizes. PDF of the original data over the whole region is plotted for reference and $\sigma(u)$ is the standard deviation of the original data.}
	\label{fig:PDF-sq_gap}
\end{figure}

To quantify the statistical similarity between the predictions and the ground truth, we can study the JS divergence, $\mathrm{JSD}(u_G)=\mathrm{JSD}(\mathrm{PDF}(u_G^{(t)})\parallel\mathrm{PDF}(u_G^{(p)}))$, defined on the distribution of the velocity amplitude in one point, which is a marginal distribution of the whole PDF of the real or predicted fields inside the gap, $p_t$ or $p_p$.
For distributions $P$ and $Q$ of a continuous random variable $x$, the JS divergence is a measure of their similarity,
\begin{equation}\label{equ:JSD}
	\mathrm{JSD}(P\parallel Q)=\frac{1}{2}\mathrm{KL}(P\parallel M)+\frac{1}{2}\mathrm{KL}(Q\parallel M),
\end{equation}
where $M=\frac{1}{2}(P+Q)$ and
\begin{equation}
	\mathrm{KL}(P\parallel Q)=\int_{-\infty}^\infty P(x)\log\left(\frac{P(x)}{Q(x)}\right)\,\mathrm{d}x
\end{equation}
is the Kullback-Leibler divergence. A small JS divergence indicates that the two probability distributions are close and vice versa. We use the base 2 logarithm and thus $0\le\mathrm{JSD}(P\parallel Q)\le1$
, with $\mathrm{JSD}(P\parallel Q)=0$ if and only if $P=Q$. 
Similar to the MSE, the JS divergence is calculated using batches of data and 10 different batches are used to obtain its mean and range of variation. The batch size used to evaluate the JS divergence is now set at 2048, which is larger than that used for the MSE, in order to improve the estimation of the probability distributions. 
Figure \ref{fig:JSD-gap_sizes-sq_gap} shows $\mathrm{JSD}(u_G)$ for the three reconstruction tools.
We have found that GAN gives smaller $\mathrm{JSD}(u_G)$ than GPOD and EPOD by an order of magnitude over almost the full range of gap sizes, indicating that the PDF of GAN prediction has a better correspondence to the ground truth. This is further shown in figure \ref{fig:PDF-sq_gap} where we present the PDFs of the predicted velocity module for different gap sizes compared with that of the original data. Besides the imprecise PDF shapes of GPOD and EPOD, we note that they are also predicting some negative values, which is unphysical for a velocity module. This problem is avoided in the GAN reconstruction, as a ReLU activation function has been used in the last layer of the generator.

\subsection{Multi-scale information}
\begin{figure}
	\centering
	\includegraphics[width=0.7\linewidth]{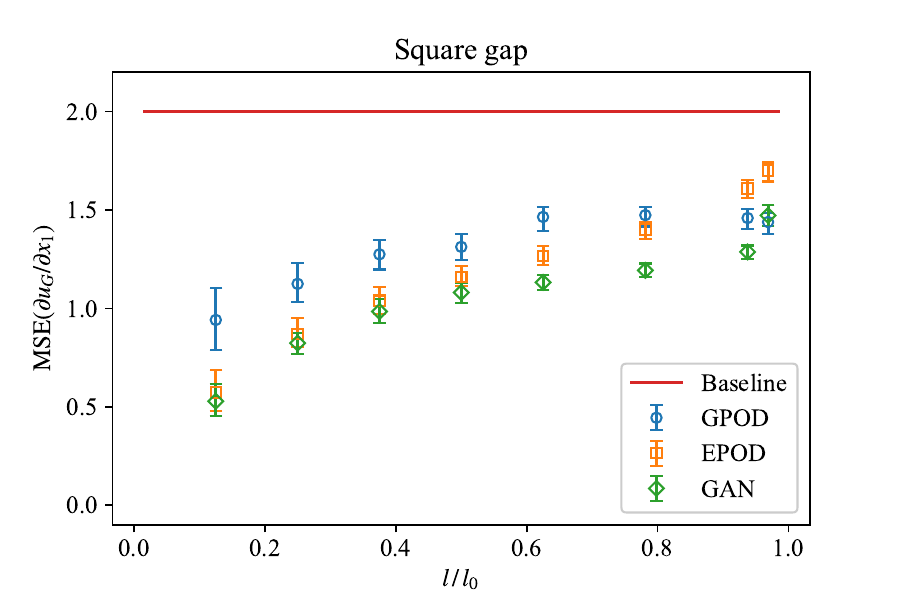}
	\caption{The MSE of the gradient of the reconstructed velocity module from GPOD, EPOD and GAN in a square gap with different sizes. Red horizontal line represents the uncorrelated baseline.}
	\label{fig:MSE-gradient-gap_sizes-sq_gap}
\end{figure}

\begin{figure}
	\hspace{0.07\linewidth}
	\includegraphics[width=0.9\linewidth]{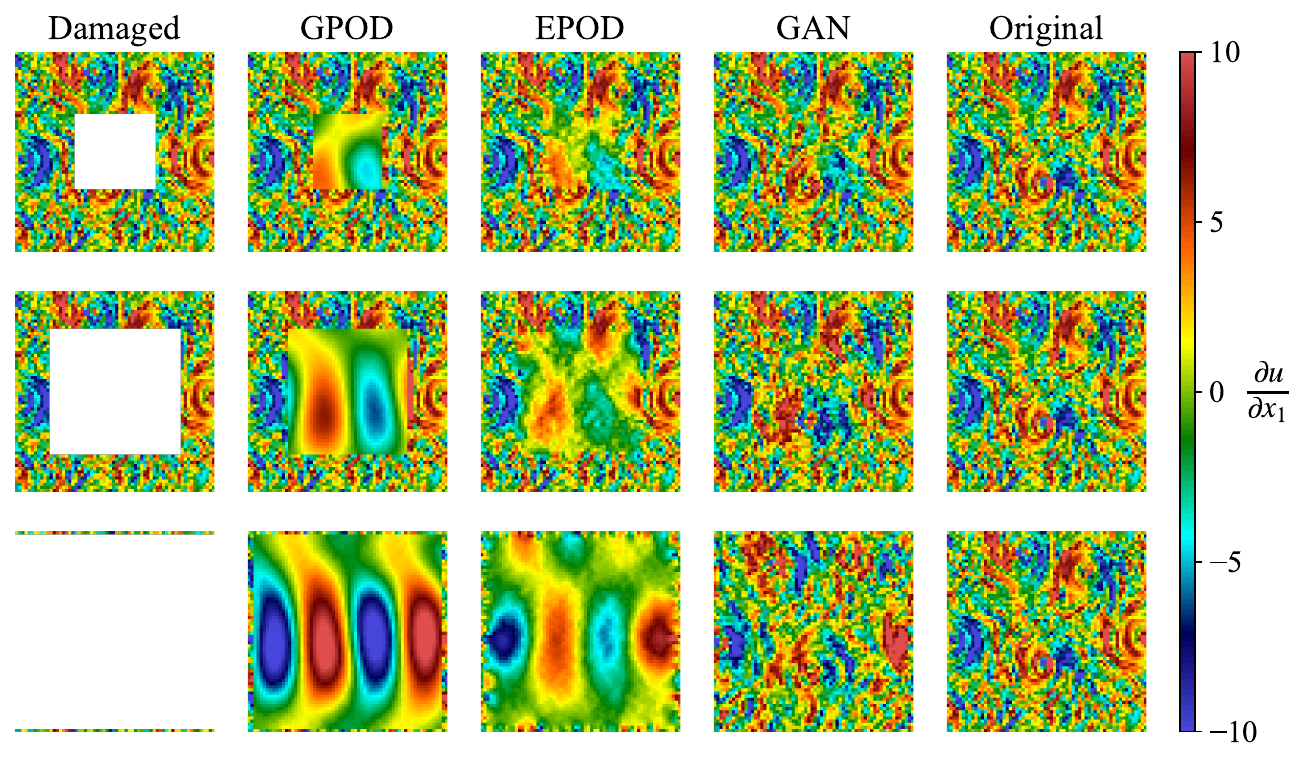}
	\caption{The gradient of the velocity module fields shown in figure \ref{fig:Reconstruction-sq_gap}. The 1st column shows the damaged fields with a square gap of sizes $l/l_0=24/64$ (1st row), $40/64$ (2nd row) and $62/64$ (3rd row). Note that for the maximum gap size, $l/l_0=62/64$, we have only one velocity layer on the vertical borders where we do not supply any information on the gradient. The 2nd to 5th columns plot the gradient fields obtained from GPOD, EPOD, GAN and the ground truth.}
	\label{fig:Reconstruction-gradient-sq_gap}
\end{figure}

\begin{figure}
	\centering
	\includegraphics[width=0.7\linewidth]{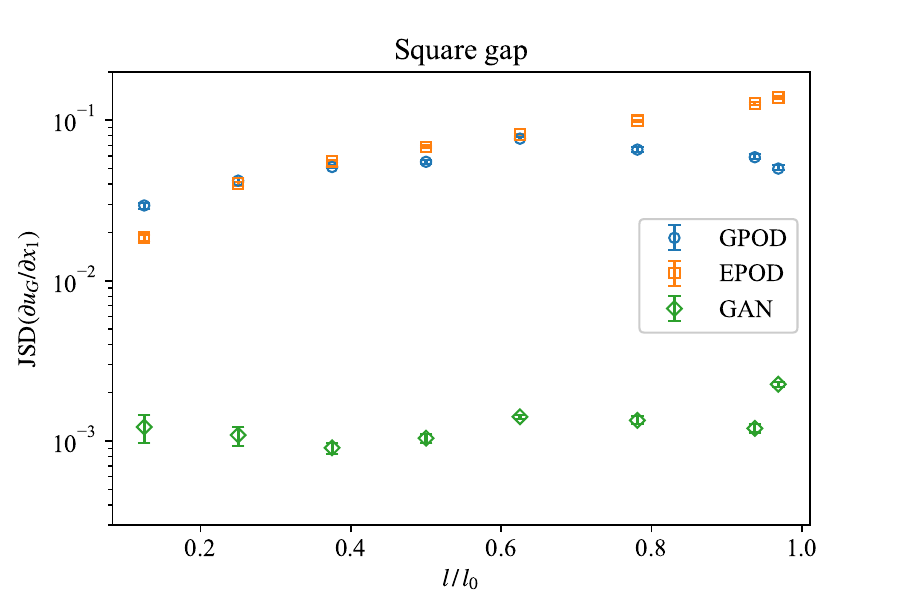}
	\caption{The JS divergence between PDFs of the gradient of the reconstructed velocity module inside the missing region from the original data and the predictions obtained from GPOD, EPOD and GAN for a square gap with different sizes.}
	\label{fig:JSD-gradient-gap_sizes-sq_gap}
\end{figure}

\begin{figure}
	\centering
	\includegraphics[width=1.0\linewidth]{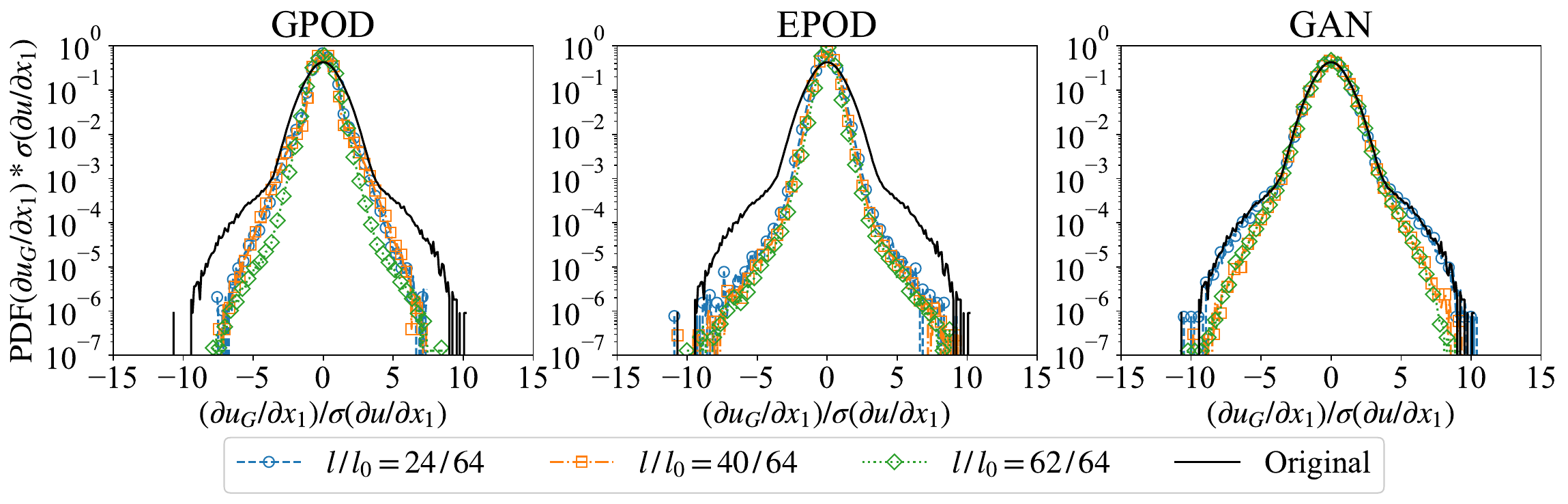}
	\caption{PDFs of the gradient of the reconstructed velocity module in the missing region obtained from GPOD, EPOD and GAN for a square gap with different sizes. PDF of the original data over the whole region is plotted for reference and $\sigma(\p u/\p x_1)$ is the standard deviation of the original data.}
	\label{fig:PDF-gradient-sq_gap}
\end{figure}

\begin{figure}
	\centering
	\includegraphics[width=1.0\linewidth]{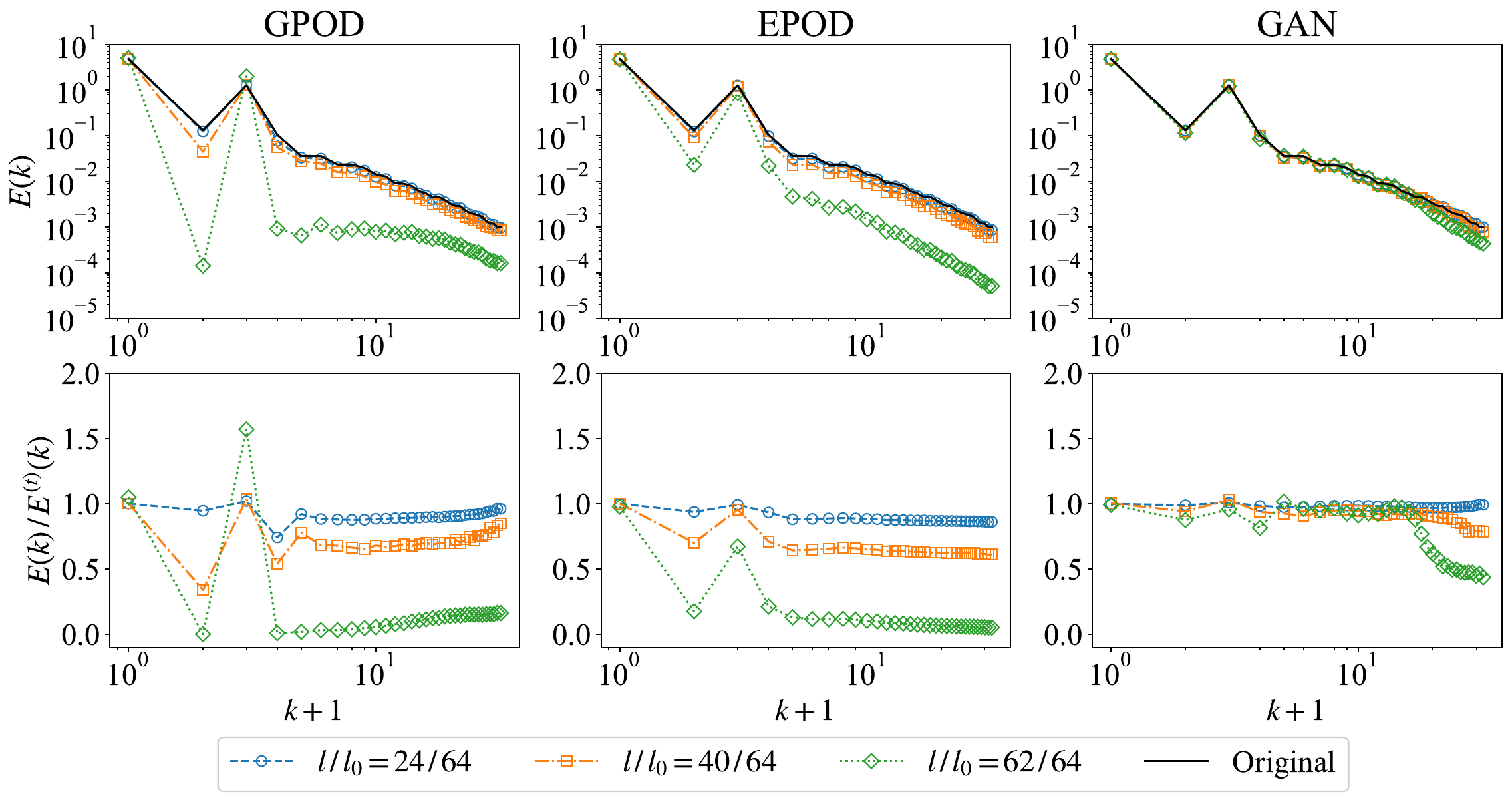}
	\caption{Energy spectra of the original velocity module and the reconstructions obtained from GPOD, EPOD and GAN for a square gap of different sizes (1st row). The corresponding $E(k)/E^{(t)}(k)$ is shown on the 2nd row, where $E(k)$ and $E^{(t)}(k)$ are the spectra of the reconstructions and the ground truth, respectively.}
	\label{fig:Spectrum-sq_gap}
\end{figure}

\begin{figure}
	\centering
	\includegraphics[width=1.0\linewidth]{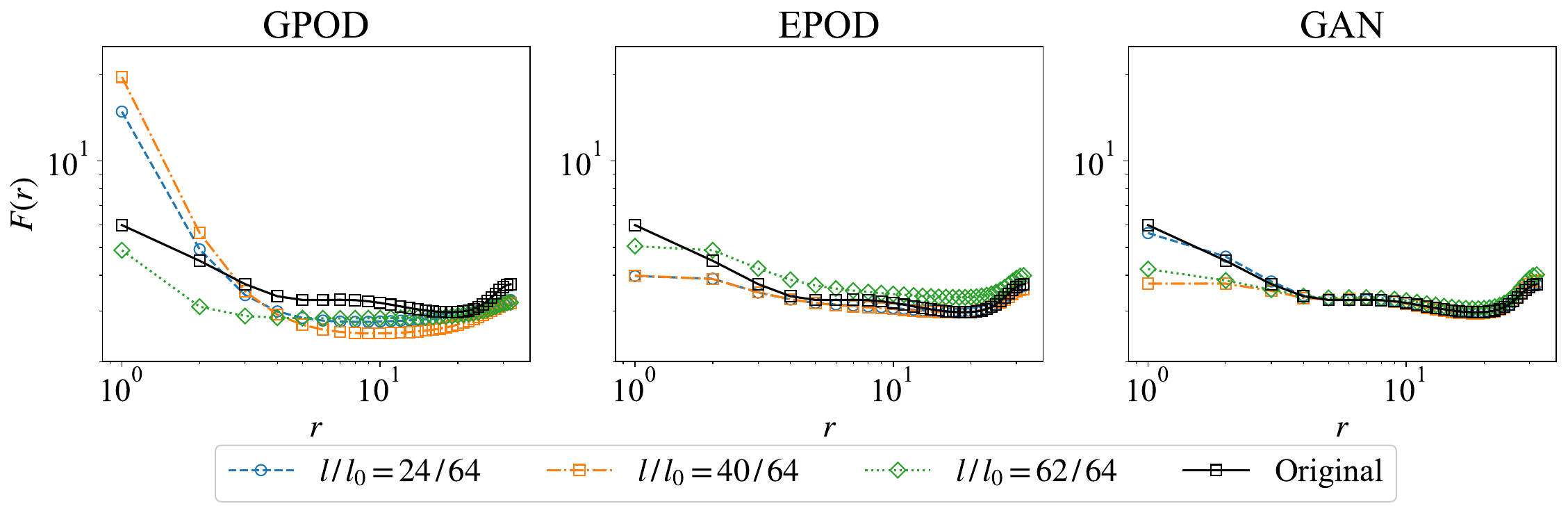}
	\caption{The flatness of the original field and the reconstructions obtained from GPOD, EPOD and GAN for a square gap of different sizes.}
	\label{fig:Flatness-sq_gap}
\end{figure}

This section reports a quantitative analysis of the multi-scale information reconstructed by the three methods. We first study the gradient of the predicted velocity module in the missing region, $\p u_G/\p x_1$.
Figure \ref{fig:MSE-gradient-gap_sizes-sq_gap} plots $\mathrm{MSE}(\p u_G/\p x_1)$, which is similarly defined as (\ref{equ:MSE}), and we can see that all methods produce $\mathrm{MSE}(\p u_G/\p x_1)$ with values much larger than those of $\mathrm{MSE}(u_G)$. Moreover, GAN shows similar errors with GPOD at the largest gap size and with EPOD at small gap sizes.
However, MSE itself is not enough for a comprehensive evaluation of the reconstruction. This can be easily understood again by looking at the gradient of different reconstructions shown in figure \ref{fig:Reconstruction-gradient-sq_gap}.
It is obvious that GAN predictions are much more `realistic' than those of GPOD and EPOD, although their values of $\mathrm{MSE}(\p u_G/\p x_1)$ are close. 
Indeed, if both fields are highly fluctuating, even a small spatial shift between the reconstruction and the true solution would result in a significantly larger MSE. 
This is exactly the case of GAN predictions where we can see that they have obvious correlations with the ground truth but the MSE is large because of its sensitivity to small spatial shifting.
On the other hand, the GPOD or EPOD solutions are inaccurate, having too small spatial fluctuations even with a similar MSE when compared with the GAN. As done above for the velocity amplitude, here we further quantify the quality of the reconstruction by looking at the JS divergence between the two PDFs in figure \ref{fig:JSD-gradient-gap_sizes-sq_gap}. For other metrics to assess the quality of the predictions see, e.g. \citep{wang2004image, wang2005translation, https://doi.org/10.48550/arxiv.2301.07541}.
Figure \ref{fig:JSD-gradient-gap_sizes-sq_gap} confirms that GAN is able to well predict the PDF of $\p u_G/\p x_1$ while GPOD and EPOD do not have this ability. Moreover, GPOD produces comparable $\mathrm{JSD}(\p u_G/\p x_1)$ with EPOD. The above conclusions are further supported in figure \ref{fig:PDF-gradient-sq_gap}, which shows PDFs of $\p u_G/\p x_1$ from the predictions and the ground truth.
We next compare the scale-by-scale energy budget of the original and reconstructed solutions in figure \ref{fig:Spectrum-sq_gap}, with the help of the energy spectrum defined over the whole region,
\begin{equation}
	E(k)=\sum_{k\le\lVert \bm{k}\rVert<k+1}\frac{1}{2}\langle\hat{u}(\bm{k})\hat{u}^\ast(\bm{k})\rangle,
\end{equation}
where $\bm{k}=(k_1, k_2)$ is the wave number, $\hat{u}(\bm{k})$ is the Fourier transform of velocity module and $\hat{u}^\ast(\bm{k})$ is its complex conjugate.
To highlight the reconstruction performance as a function of the wave number, we also show the ratio between the reconstructed and the original spectra, $E(k)/E^{(t)}(k)$ for the three different gap sizes on the second row of figure \ref{fig:Spectrum-sq_gap}. 
Figure \ref{fig:Flatness-sq_gap} plots the flatness of the reconstructed fields, 
\begin{equation}\label{equ:flatness}
	F(r)=\langle(\delta_r u)^4\rangle/\langle(\delta_r u)^2\rangle,
\end{equation}
where $\delta_r u=u(\bm{x}+\bm{r})-u(\bm{x})$ and $\bm{r}=(r, 0)$. The angle bracket in (\ref{equ:flatness}) represents the average over the test dataset and over $\bm{x}$, for which $\bm{x}$ or $\bm{x}+\bm{r}$ fall in the gap. The flatness of the ground truth calculated over the whole region is also shown for reference.
We remark that the flatness is used to characterize the intermittency in the turbulence community \citep[see][]{frisch1995turbulence}. It is determined by the two-point PDFs, $\mathrm{PDF}(\delta_ru)$, connected to the distribution of the whole real or generated fields inside the gap, $p_t$ or $p_p$. 
Figures \ref{fig:Spectrum-sq_gap} and \ref{fig:Flatness-sq_gap} show that GAN performs well to reproduce the multi-scale statistical properties, except at small scales for large gap sizes. However, GPOD and EPOD can only predict a good energy spectrum for the small gap size $l/l_0=24/64$ but fail at all scales for both the energy spectrum and flatness at gap sizes $l/l_0=40/64$ and $62/60$.

\subsection{Extreme events}
\begin{figure}
	\centering
	\includegraphics[width=1.0\linewidth]{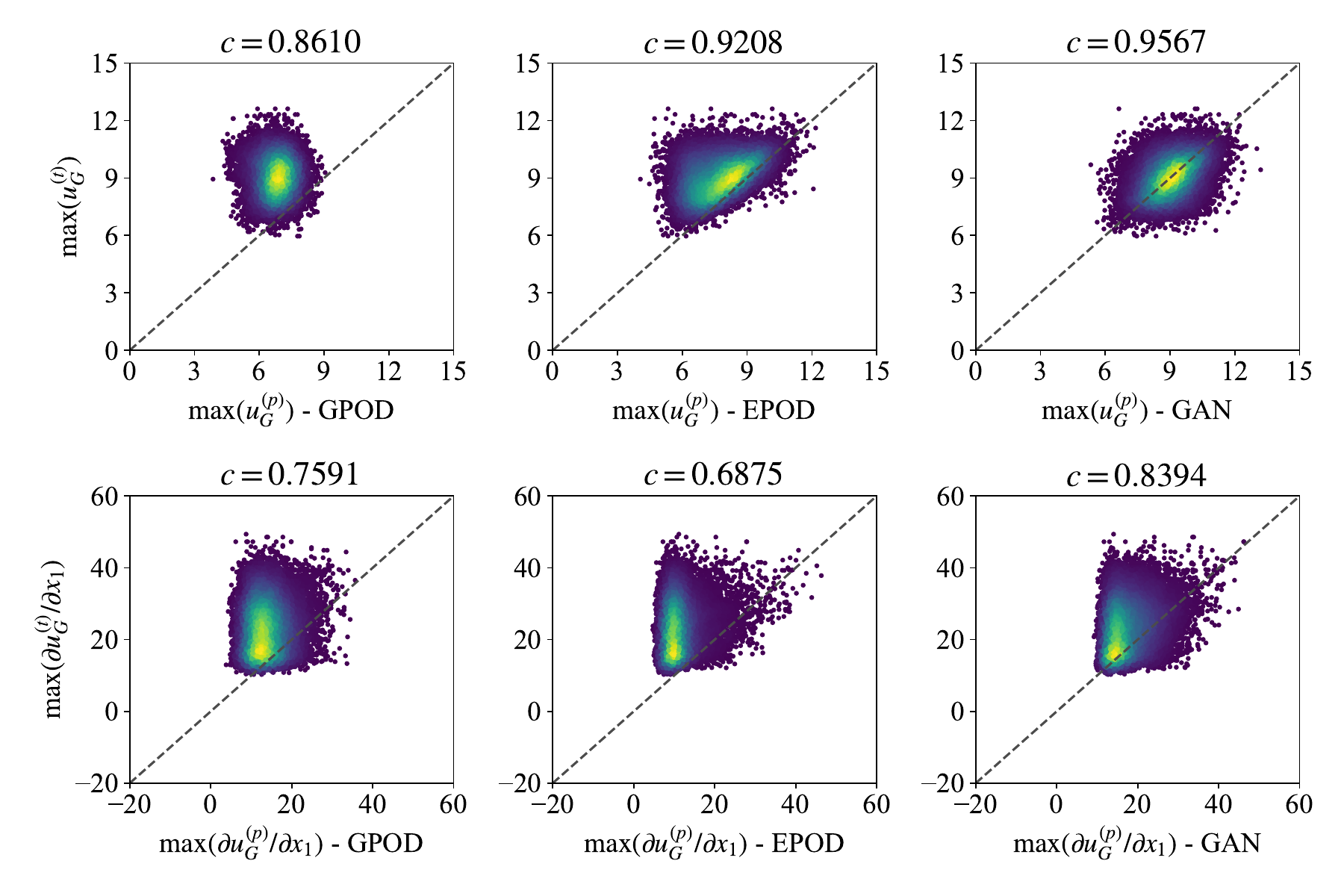}
	\caption{Scatter plots of the maximum values of velocity module (1st row) or its gradient (2nd row) in the missing region obtained from the original data and the one produced by GPOD, EPOD or GAN for a square gap of size $l/l_0=40/64$. Colors are proportional to the density of points in the scatter plot. The correlation indices are shown on top of each panel.}
	\label{fig:Extreme_values-sq_gap40}
\end{figure}

In this section, we focus on the ability of the different methods to reconstruct extreme events inside the gap for each frame. In figure \ref{fig:Extreme_values-sq_gap40} we present the scatter plots of the largest values of velocity module or its gradient measured in the gap region from the original data and the predicted fields generated by GPOD, EPOD or GAN. On top of each panel we report the scatter plot correlation index, defined as+
\begin{equation}
	c=\langle1-|\sin\theta|\rangle,
\end{equation}
where $|\sin\theta|=\lVert\bm{U}\times\bm{e}\rVert/\lVert\bm{U}\rVert$ with $\theta$ as the angle between the unit vector $\bm{e}=(1/\sqrt{2},1/\sqrt{2})$ and $\bm{U}=(\max(u_G^{(p)}),\max(u_G^{(t)}))$. The $\bm{U}$ for $\p u_G/\p x_1$ can be similarly defined.
It is obvious that $c\in[0,1]$ and $c=1$ corresponds to a perfect prediction in terms of the extreme events. In figure \ref{fig:Extreme_values-sq_gap40}, it shows that for both extreme values of velocity module and its gradient, GAN is the least biased while the other two methods tend to underestimate them.

\section{Dependency of GAN-based reconstruction on the adversarial ratio}
\label{sec:depen}
\begin{figure}
	\centering
	\includegraphics[width=0.7\linewidth]{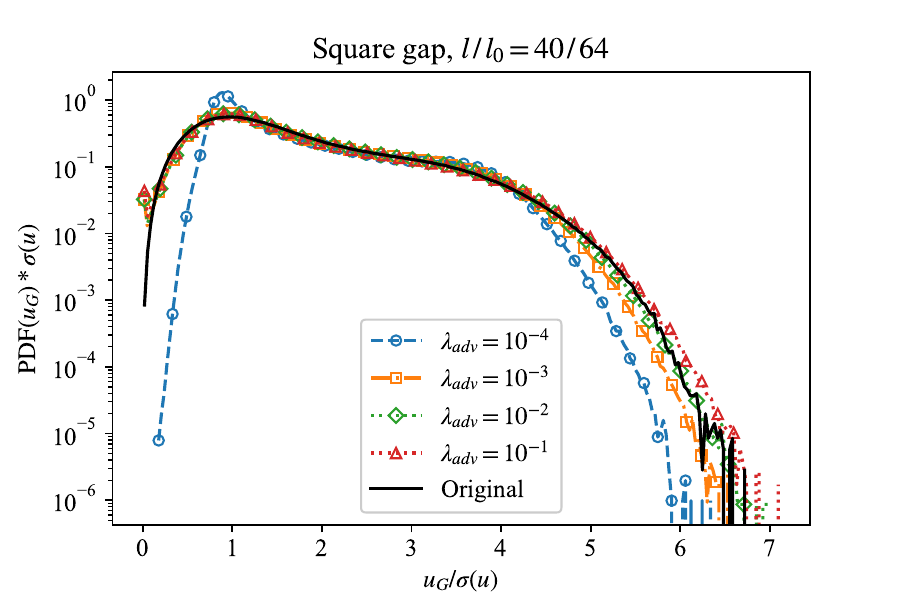}
	\caption{PDFs of the reconstructed velocity module inside the gap region, which is obtained from GAN with different adversatial ratios, for a square gap of size $l/l_0=40/64$.}
	\label{fig:PDF-ARs-sq_gap40}
\end{figure}

As shown by the previous results, GAN is certainly superior regarding metrics evaluated in this study. This supremacy is given by the fact that with the non-linear CNN structure of the generator, GAN optimizes the point-wise $L_2$ loss and minimizes the JS divergence between the probability distributions of the real and generated fields with the help of the adversarial discriminator (see \S\ref{subsec:ganba}).
To study the effects of the balancing between the above two objectives on reconstruction quality,
we have performed a systematic scanning of the GAN performances at changing the adversarial ratio $\lambda_{adv}$, the hyper-parameter controlling the relative importance of $L_2$ loss and adversarial loss of the generator, as shown in equation (\ref{equ:L_GAN}). 
We consider a central square gap of size $l/l_0=40/64$ and train the GAN with different adversarial ratios, where $\lambda_{adv}=10^{-4}$, $10^{-3}$, $10^{-2}$ and $10^{-1}$. 
Table \ref{tab:MSE-JSD-ARs-sq_gap40} shows the values of $\mathrm{MSE}(u_G)$ and $\mathrm{JSD}(u_G)$ obtained at different adversarial ratios.
\begin{table}
	\begin{center}
		\def~{\hphantom{0}}
		\begin{tabular}{ccccc|cc}
			$\lambda_{adv}$ & $10^{-4}$ & $10^{-3}$ & $10^{-2}$ & $10^{-1}$ & GPOD & EPOD\\[5pt]
			$\mathrm{MSE}(u_G)(\times10^2)$ & $6.2_{-0.7}^{+0.8}$ & $7.8_{-0.7}^{+0.6}$ & $8.3_{-0.6}^{+0.7}$ & $9.0_{-0.8}^{+0.7}$ & $15_{-1}^{+1}$ & $9.2_{-0.7}^{+0.7}$\\[5pt]
			$\mathrm{JSD}(u_G)(\times10^3)$ & $65_{-1}^{+1}$ & $2.6_{-0.1}^{+0.2}$ & $2.1_{-0.1}^{+0.1}$ & $2.06_{-0.11}^{+0.05}$ & $43_{-1}^{+1}$ & $18_{-1}^{+1}$\\
		\end{tabular}
		\caption{The MSE and the JS divergence between PDFs for the original and generated velocity module inside the missing region, obtained from GAN with different adversarial ratios for a square gap of size \(l/l_0=40/64\). The results for GPOD and EPOD are provided as well for comparison. The MSE and JS divergence are computed over different test batches, specifically of sizes 128 and 2048, respectively. From these computations, we obtain both the mean values and the error bounds.}
		\label{tab:MSE-JSD-ARs-sq_gap40}
	\end{center}
\end{table}
It is obvious that the adversarial ratio controls the balance between the point-wise reconstruction error and the predicted turbulent statistics. As the adversarial ratio increases, the MSE increases while the JS divergence decreases. PDFs of the predicted velocity module from GANs with different adversarial ratios are compared with that of the original data in figure \ref{fig:PDF-ARs-sq_gap40}, which shows that the predicted PDF gets closer to the original one with a larger adversarial ratio.
The above results clearly show that there exists an {\it optimal} adversarial ratio to satisfy the multi-objective requirements of having a small $L_2$ distance and a realistic PDF. In the limit of vanishing $\lambda_{adv}$, the GAN outperforms GPOD and EPOD in terms of MSE, but falls behind them concerning JS divergence (table \ref{tab:MSE-JSD-ARs-sq_gap40}).

\section{Dependency on gap geometry: random gappiness}
\label{sec:rando}
\begin{figure}
	\centering
	\includegraphics[width=1.0\linewidth]{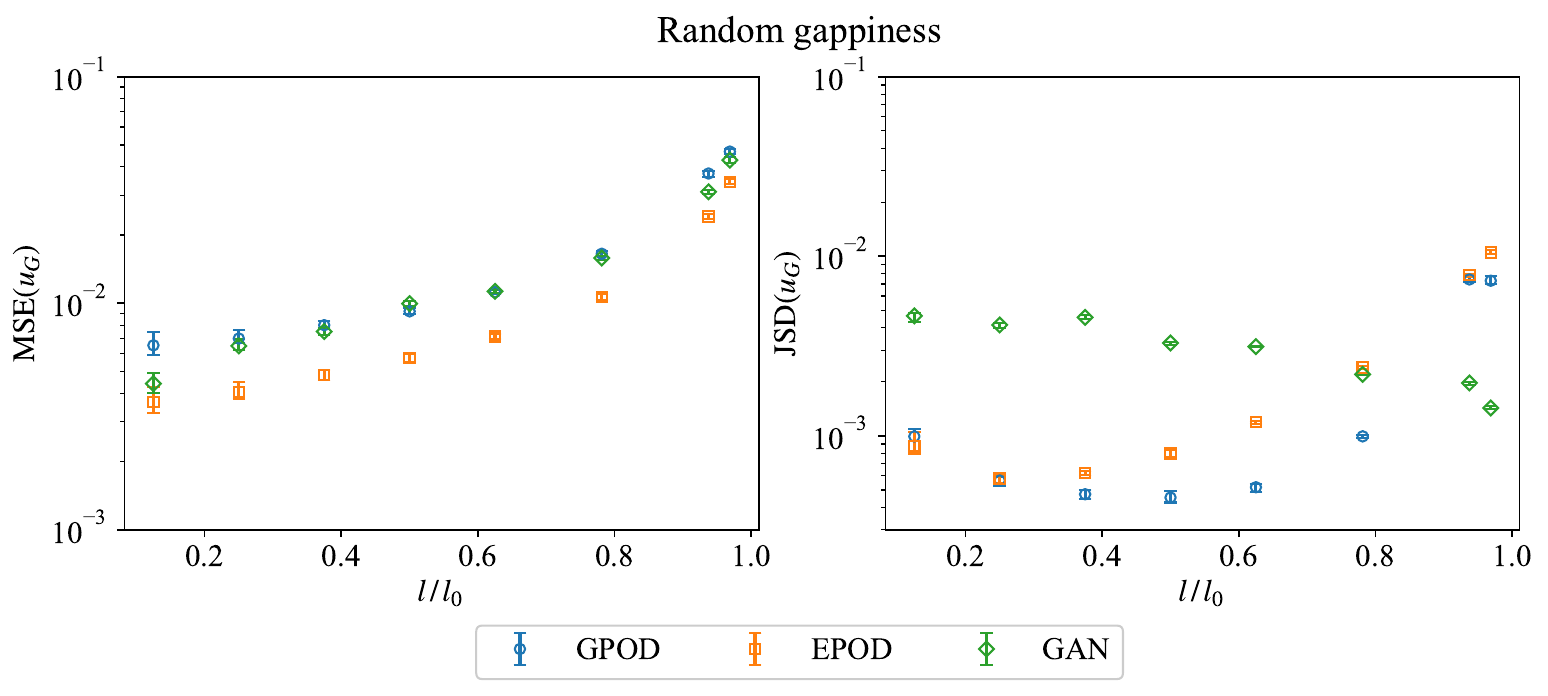}
	\caption{The MSE (left) and the JS divergence (right) between PDFs for the original and generated velocity module inside the missing region, obtained from GPOD, EPOD and GAN for random gappiness with different sizes.}
	\label{fig:MSE-gap_sizes-rp_gap}
\end{figure}

\begin{figure}
	\centering
	\includegraphics[width=1.0\linewidth]{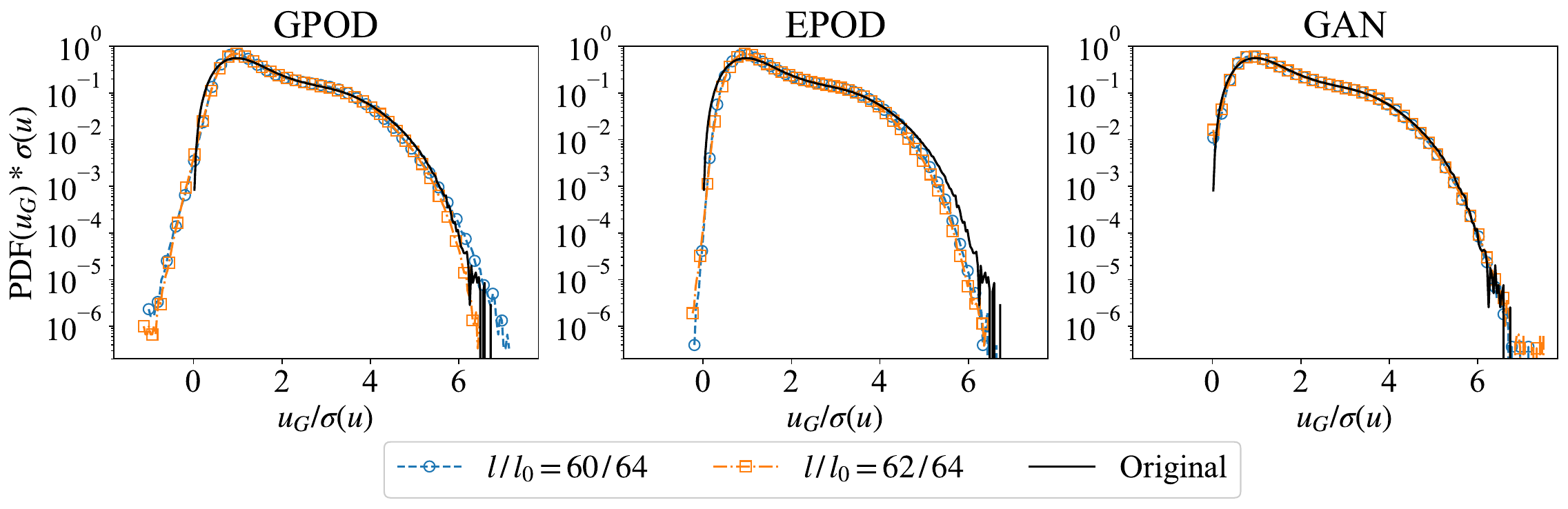}
	\caption{PDFs of the velocity module in the missing region obtained from GPOD, EPOD and GAN for random gappiness with different sizes. PDF of the original data over the whole region is plotted for reference and $\sigma(u)$ is the standard deviation of the original data.}
	\label{fig:PDF-rp_gap}
\end{figure}

\begin{figure}
	\centering
	\includegraphics[width=1.0\linewidth]{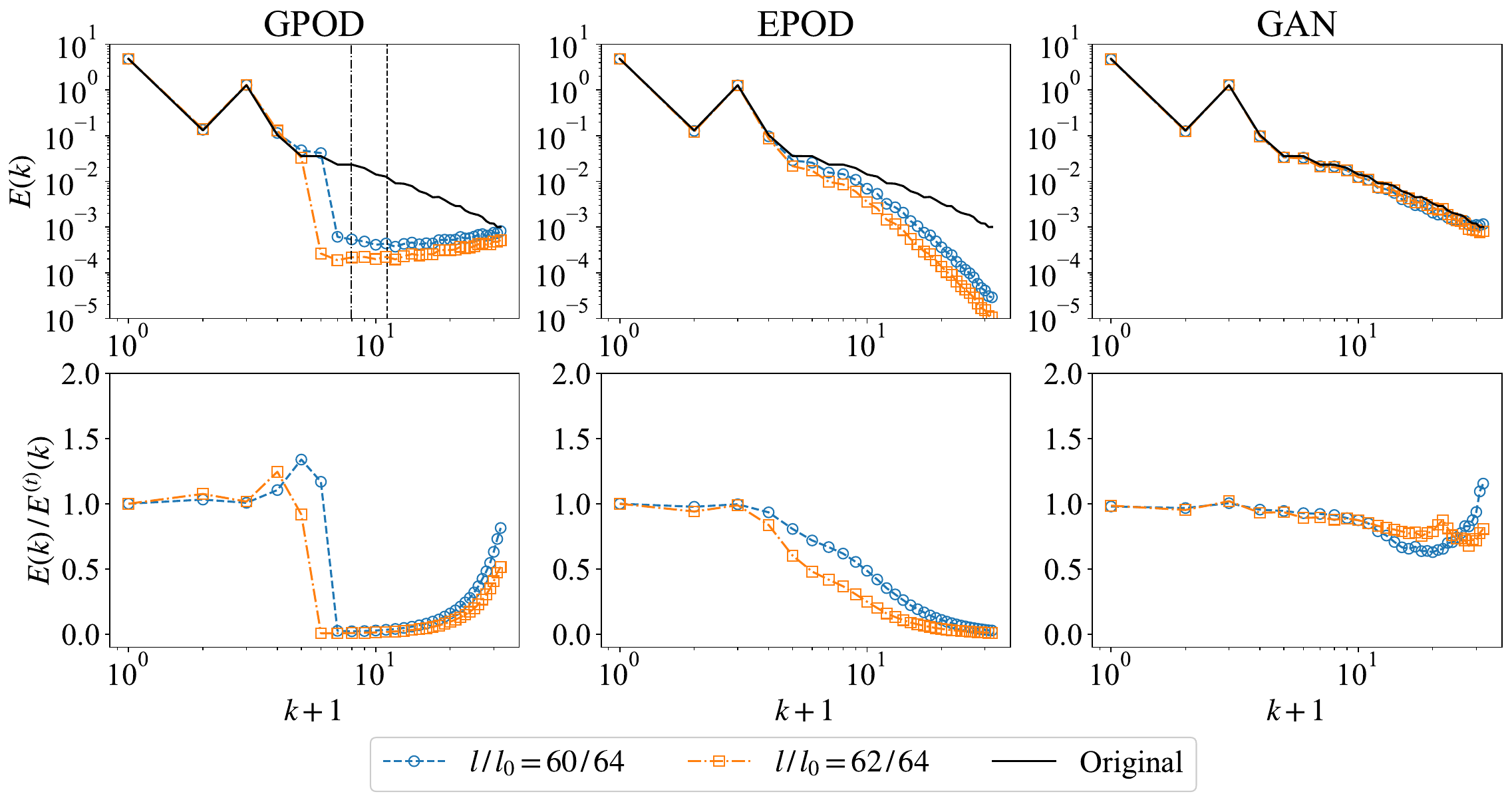}
	\caption{Energy spectra of the original velocity module and the reconstructions obtained from GPOD, EPOD and GAN for random gappiness with different sizes (1st row). The corresponding $E(k)/E^{(t)}(k)$ is shown on the 2nd row, where $E(k)$ and $E^{(t)}(k)$ are the spectra of the reconstructions and the ground truth, respectively. The vertical dashed and dash-dot lines respectively indicate the wave numbers corresponding to the data resolutions $l/l_0=60/64$ and $62/64$. These wave numbers are calculated as $k_\eta/d$, where $d$ is the corresponding downsampling rate.}
	\label{fig:Spectrum-rp_gap}
\end{figure}

\begin{figure}
	\hspace{0.07\linewidth}
	\includegraphics[width=0.9\linewidth]{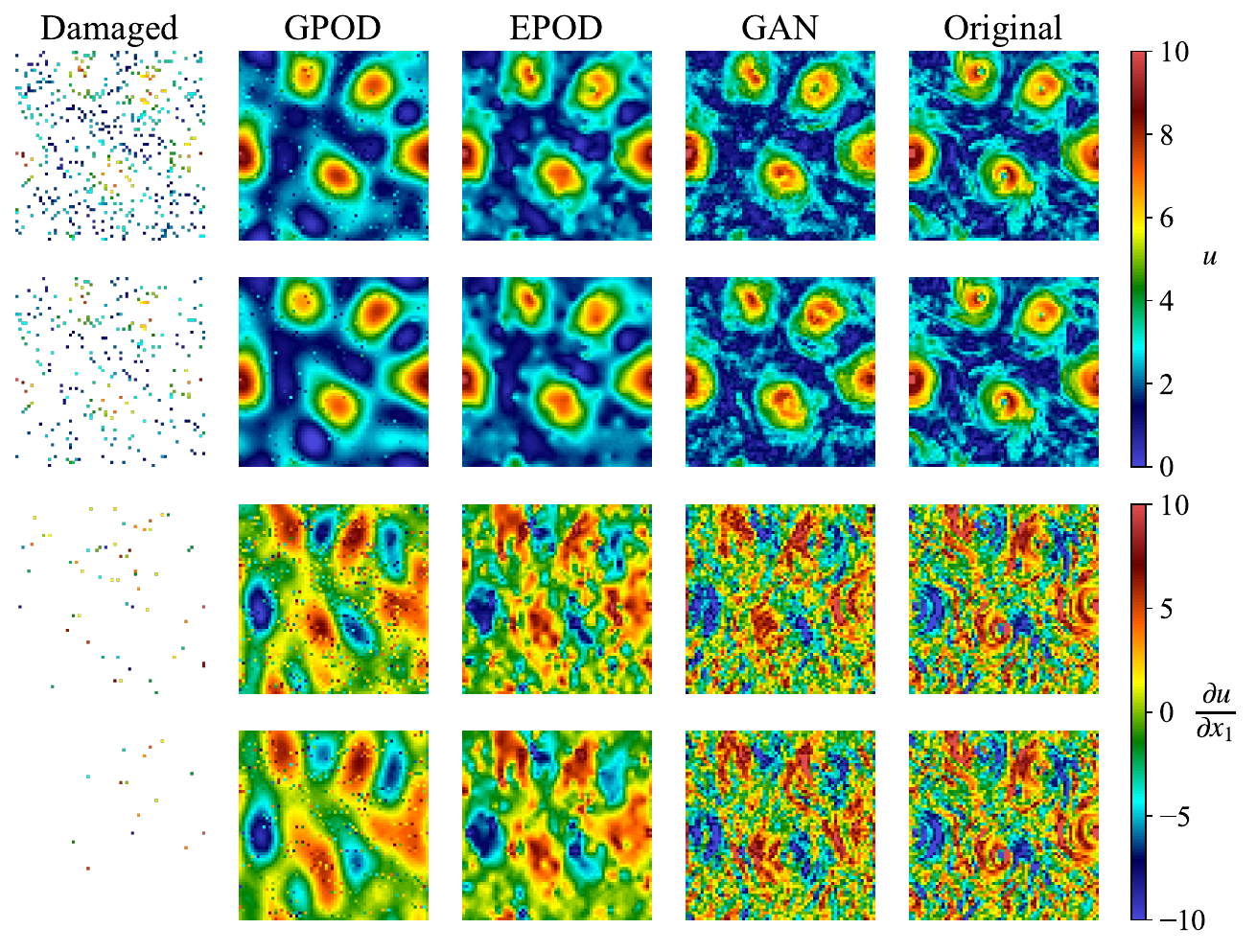}
	\caption{Reconstruction of an instantaneous field (velocity module) by the different tools for random gappiness of sizes $l/l_0=60/64$ (1st row) and $l/l_0=62/64$ (2nd row). The corresponding gradient fields are shown in the 3rd and 4th rows. The damaged fields are shown in the 1st column, while the 2th to 4th columns show the fields obtained from GPOD, EPOD and GAN. The ground truth is shown in the 5th column.}
	\label{fig:Reconstruction-rp_gap}
\end{figure}

Things change when looking at a completely different topology of the {\it damages}. Here we study the case ii) in \S\ref{subsec:dataset}, where position points are removed randomly in the original domain $I$, without any spatial correlations. 
Because the random gappiness is easier for interpolating than a square gap of the same size, all reconstruction methods show good and comparable results in terms of the MSE, the JS divergence and PDFs for velocity module (figures \ref{fig:MSE-gap_sizes-rp_gap} and \ref{fig:PDF-rp_gap}).
For almost all damaged densities, POD- and GAN-based methods give small values of $\mathrm{MSE}(u_G)$ and $\mathrm{JSD}(u_G)$. However, when the total damaged region area is extremely large, GPOD and EPOD are not able to reconstruct the field at large wave numbers while GAN still works well because of the adversarial training, as shown by the energy spectra in figure \ref{fig:Spectrum-rp_gap}. 
Figure \ref{fig:Reconstruction-rp_gap} shows the reconstruction of the velocity module and the corresponding gradient fields for random gappiness with two extremely large sizes.
It is obvious that GPOD and EPOD only predict the large-scale structure while GAN generates reconstructions with multi-scale information. To make a comparison between the random gappiness case and the super-resolution task, we can compare the effect of random gappiness of sizes $l/l_0=60/64$ and $62/64$, similar to a downsampling of the original field by approximately a factor of $3$ and $4$ for each spatial direction, respectively.

\section{Dependency on measurement noise and computational costs}\label{sec:discu}
\begin{figure}
	\centering
	\includegraphics[width=1.0\linewidth]{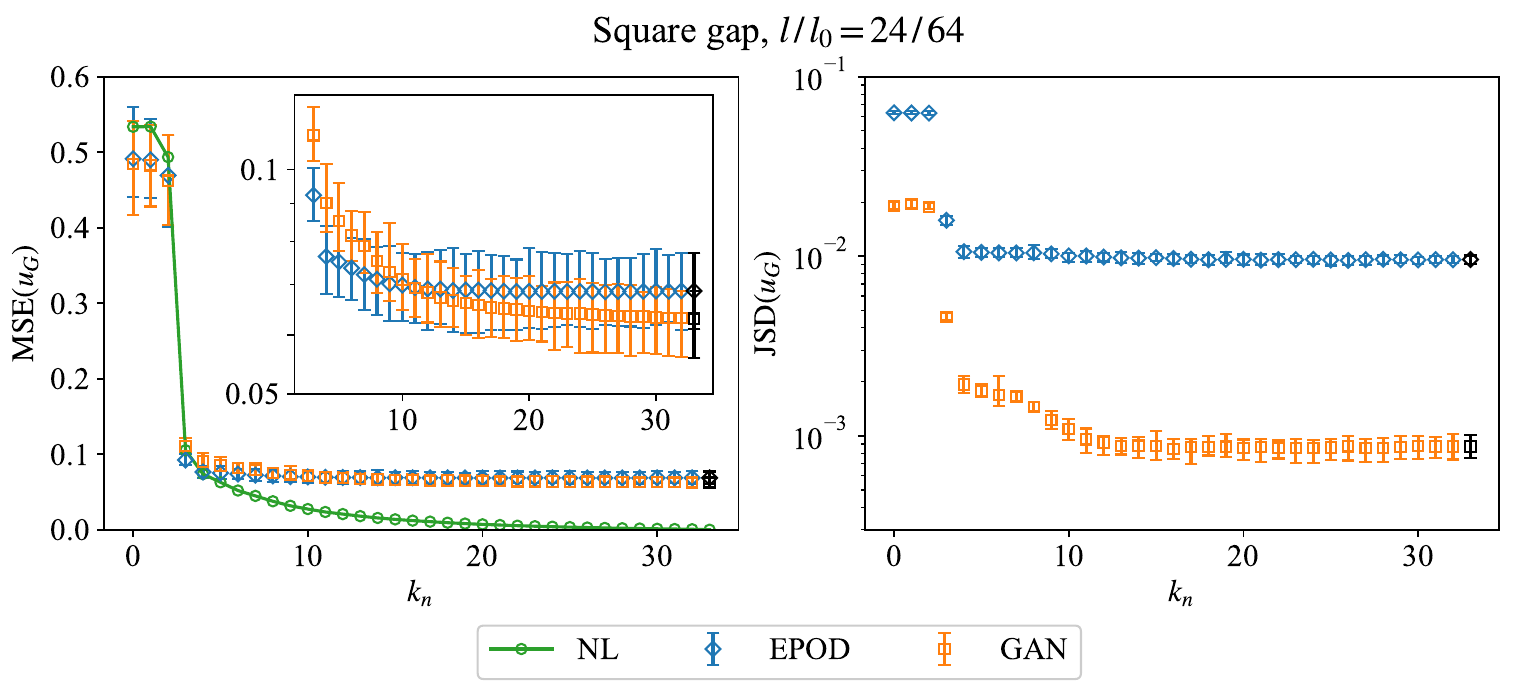}
	\caption{The MSE (left) and the JS divergence (right) between PDFs for the original and generated velocity module inside the missing region, obtained from EPOD and GAN with input measurements of different noise levels for a square gap of size $l/l_0=24/64$. The results obtained with the noiseless input are plotted with black symbols at the right end of each panel. Results obtained with noiseless input are represented by black symbols, positioned at the right end of each panel. The estimate of the noise level introduced in the physical space is given by the green curve (NL). The inset box presents the MSE on a log-lin scale.}
	\label{fig:EPOD-GAN-norm_MSE-JSD-sq_gap24-noisy_kns}
\end{figure}

\begin{figure}
	\centering
	\includegraphics[width=1.0\linewidth]{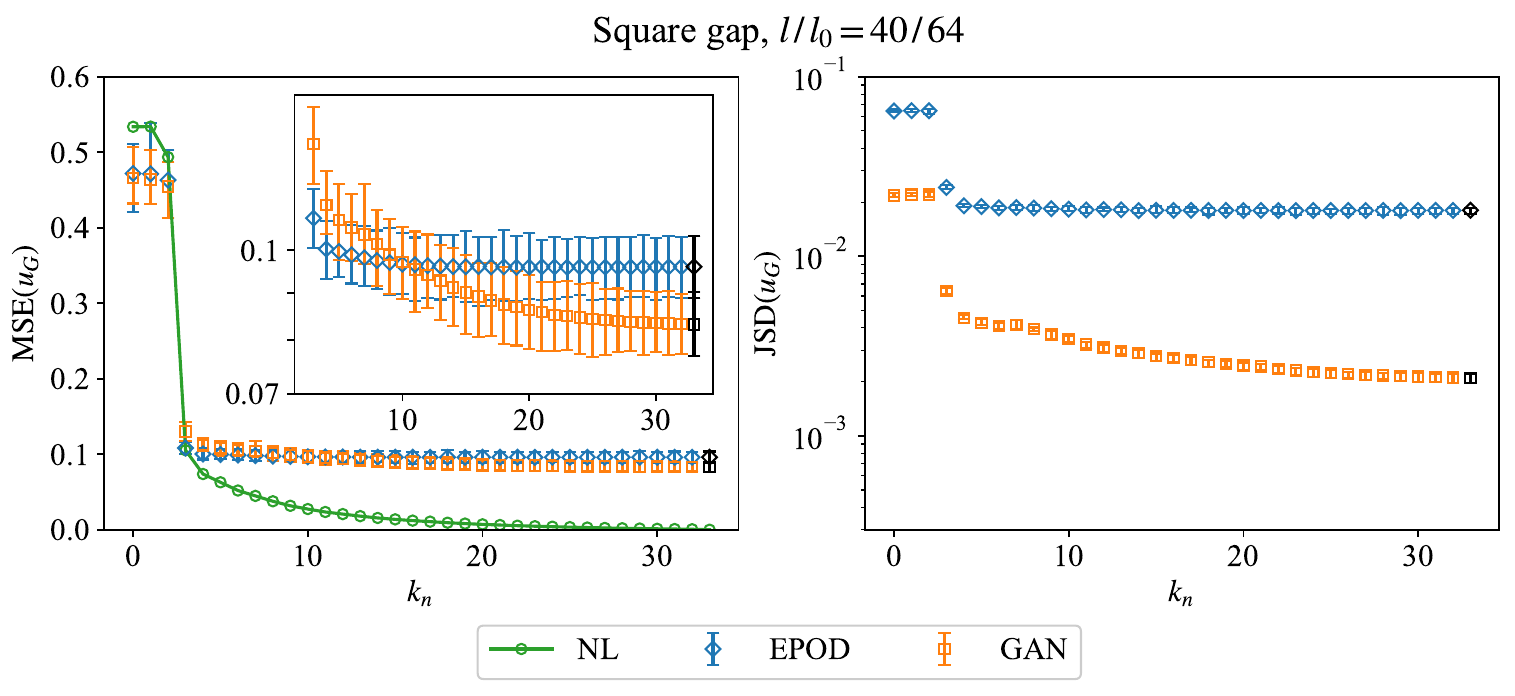}
	\caption{The same as figure \ref{fig:EPOD-GAN-norm_MSE-JSD-sq_gap24-noisy_kns} but for a square gap of size $l/l_0=40/64$.}
	\label{fig:EPOD-GAN-norm_MSE-JSD-sq_gap40-noisy_kns}
\end{figure}

So far we have investigated the reconstruction of turbulent data without noise and therein the measurement resolution equals to the Kolmogorov scale, as shown in (\ref{equ:filter}). However, field measurements are usually noisy. The noise can come from the errors encountered experimentally and/or a lack of resolution, such as for the filtered data in PIV. 
In this section, we evaluate the robustness of EPOD and GAN methods by considering a scenario where the magnitude of the velocity module remains unchanged, while its phase is randomly perturbed for wave numbers above a threshold value, $k_n$.
We estimate the noise level in the physical space, represented as $\mathrm{NL}(k_n)$, as the MSE of the noisy data with respect to the original fields. Contrary to equation (\ref{equ:MSE}), which is averaged over the gap region $G$, the noise level in this case is averaged across the entire domain $I$. Given the noise properties, we have $\mathrm{NL}(k_n)\sim\sum_{\lVert \bm{k}\rVert\geq k_n}E(k)$ (see figure \ref{fig:Dataset}(b)).
The noisy measurements in the known region $S$ are then fed into the reconstruction models, which have already been trained with the noiseless data. It is important to remark that the predictions are evaluated with the ground truth with no noise.  
Figures \ref{fig:EPOD-GAN-norm_MSE-JSD-sq_gap24-noisy_kns} and \ref{fig:EPOD-GAN-norm_MSE-JSD-sq_gap40-noisy_kns} show the $\mathrm{MSE}(u_G)$ and $\mathrm{JSD}(u_G)$ obtained from EPOD and GAN with input measurements of different noise levels for a square gap of sizes $l/l_0=24/64$ and $40/64$, respectively. The results obtained with the noiseless input are also shown with black symbols at the right end of each panel. Both MSE and JS divergence of the velocity module are more sensitive to the most energetic large-scale properties, while they change slightly when the noise is applied at $k_n\ge3$, indicating the robustness of both approaches for these cases. EPOD and GAN predict drastically worse results when the noise is applied at $k_n\le2$.

Concerning the computational cost of the three methods here studied, we remark that during the training process of GPOD one first conducts a singular value decomposition (SVD) to solve (\ref{equ:eig_prob1}), with a computational cost $\sim \mathcal{O}(N_\mathrm{train}^2N_I)$. This operation is followed by an optimization of $N'$ by scanning all possible values ($ 1 \le N' \le N_S$). This second step adds an extra computational cost of $\mathcal{O}(N_S^2N_IN_\mathrm{train})$ for the required linear algebra operations as discussed in Appendix \ref{appA}, see (\ref{equ:err_gpod}), (\ref{equ:coef_b}), (\ref{equ:rec_field}) and (\ref{equ:rec_err1}). The GPOD testing process is conducted according to (\ref{equ:coef_b}) and (\ref{equ:rec_field}) with a computational cost of $\mathcal{O}(N'N_IN_\mathrm{test})$. 
The training process of EPOD is computationally cheaper than that of GPOD. The cost can be estimated as $\mathcal{O}(N_\mathrm{train}^2N_S)$ considering a SVD to solve (\ref{equ:eig_prob2}) and the linear algebra operations in (\ref{equ:coe}) and (\ref{equ:ext}). The testing process of EPOD consists of carrying out (\ref{equ:coe}) and (\ref{equ:epod}), with a computational cost of $\mathcal{O}(N_\mathrm{test}N_SN_I)$. GAN is the most computationally expensive method. It has about $5\times10^6$ ($\gg N_\mathrm{train}$) trainable parameters, which are involved in the forward and the backward propagation for all the training data in one epoch. Moreover, hundreds of epoch are required for the convergence of GAN. However, benefiting from the GPU hardware, 
GAN training requires only 4 hours on an A100 Nvidia GPU. Once trained, all methods are highly efficient in performing reconstruction. It is important to emphasize that any improvement over existing methods is valuable, regardless of the computational cost involved. Even when computational resources are not a constraint, GPOD and EPOD cannot further improve the accuracy of the reconstruction. This limitation is attributed to the linear estimation of the flow state inherent in these methods. Nevertheless, there is still potential for further improvement of the GAN results, as numerous hyper-parameters remain to be fine-tuned. These hyper-parameters include aspects such as the depth of the networks, the dimension of the latent feature, etc.

\section{Conclusions}
\label{sec:concl}

In this work, two linear POD-based approaches, GPOD and EPOD, are compared against GAN, consisting of two adversarial non-linear CNNs, to reconstruct 2D damaged fields taken from a database of 3D rotating turbulent flows. Performances have been quantitatively judged on the basis of (i) $L_2$ distance between each the ground truth and the reconstructed field, (ii) statistical validations based on JS divergence between the one-point PDFs, (iii) spectral properties and multi-scale flatness, and (iv) extreme events for a single frame. For one central square gap the GAN approach is proved to be superior to GPOD and EPOD, when both MSE and JS divergence are simultaneously considered, in particular for large gap sizes where the missing of multi-scale information makes the task extremely difficult. Moreover, GAN predictions are also better in terms of the energy spectra and flatness, as well as for the predicted extreme events. In the presence of random damages, the three approaches give similar results except for the case of extreme gappiness where GAN is leading again.


GPOD always generates `discontinuous' predictions with respect to the supplied measurements. This is because GPOD only minimizes the $L_2$ distance and the optimal number of POD modes used is usually much smaller than the number of measured points. On the other hand, EPOD considers the correlation between the fields inside and outside the gap and its predictions have a number of degrees of freedom equal to the number of measured points. Compared with GPOD, EPOD is less computationally demanding and generates better predictions. When the gap is extremely large, neither GPOD nor EPOD gives satisfying predictions as they have too few degrees of freedoms.

With the help of adversarial training, GAN can optimize a multi-objective problem, minimizing simultaneously the $L_2$ distance frame by frame and the JS divergence between the real and generated distributions of the whole fields in the missing region. Furthermore, we show that for GAN reconstructions, large adversarial ratios undermine the MSE but improve the generated statistical properties and vice versa.


In terms of the potential for practical applications of the three tools analyzed in this study, we have demonstrated that both EPOD and GAN exhibit robust properties when faced with noisy multi-scale measurements. It is also worth noting that in many applications, gaps can also arise in the Fourier space. This typically occurs when we encounter measurement noise or modeling limitations at high wave numbers. In such situations, we face a super-resolution problem where we need to reconstruct the missing small-scale information.



Our work is a first step toward the set-up of benchmarks and grand challenges for realistic turbulent problems with interest in geophysical and laboratory applications, where the lack of measurements obstructs the capability to fully control the system. Many questions remain open, connected to the performance of different GAN architectures, and the difficulty of having apriori estimates of the deepness and complexity of the GAN architecture as a function of the complexity of the physics, in particular concerning the quantity and the geometry (2D or 3D) of the missing information. Furthermore, little is known about the performance of the data-driven models as a function of the Reynolds or Rossby numbers, and the possibility to supply physics information to help to further improve the network's performances.


\bigskip
\begin{flushleft}
	{\bf Funding}. This work was supported by the European Research Council (ERC) under the European Union's Horizon 2020 research and innovation programme (grant agreement No. 882340); the NSFC (M. W., grant nos 12225204, 91752201 and 11988102); Shenzhen Science \& Technology Program (M. W., grant No. KQTD20180411143441009); and Department of Science and Technology of Guangdong Province (M. W., grant nos 2019B21203001, 2020B1212030001).
\end{flushleft}

\bigskip
\begin{flushleft}
	{\bf Declaration of Interests}. The authors report no conflict of interest.
\end{flushleft}

\appendix

\section{Error analysis of GPOD reconstruction}\label{appA}
Consider the whole region $I$, the gap region $G$ and the known region $S$ as sets of positions. 
%
Given \(m(\cdot)\) as function returning the number of elements in a set, we can define
\begin{equation}
	\begin{split}
		I &= \{ x_1, x_2, \ldots, x_{m(I)} \}, \\
		G &= \{ x_k \mid x_k \text{ in the gap} \} = \{ x_{i_1}, x_{i_2}, \ldots, x_{i_{m(G)}} \}, \\
		S &= I \setminus G = \{ x_{j_1}, x_{j_2}, \ldots, x_{j_{m(S)}} \},
	\end{split}
\end{equation}
and
\begin{equation}
	\begin{split}
		\bm{x} &=
		\begin{bmatrix}
			x_1 & x_2 & \cdots & x_{m(I)}
		\end{bmatrix}^T, \\
		\bar{\bm{x}} &=
		\begin{bmatrix}
			x_{i_1} & x_{i_2} & \cdots & x_{i_{m(G)}}
		\end{bmatrix}^T, \\
		\tilde{\bm{x}} &=
		\begin{bmatrix}
			x_{j_1} & x_{j_2} & \cdots & x_{j_{m(S)}}
		\end{bmatrix}^T.
	\end{split}
\end{equation}
With $N'$ as the number of POD modes kept for dimension reduction, the POD decomposition
\begin{equation}
	u(\bm{x}) = \sum_{n=1}^{N_I} a_n\psi_n(\bm{x}) = \sum_{n=1}^{N'} a_n\psi_n(\bm{x}) + \sum_{n=N'+1}^{N_I} a_n\psi_n(\bm{x})
\end{equation}
can be written in the vector form
\begin{equation}
	\bm{u} = \bm{Xa} = \bm{X}'\bm{a}' + \bm{r}',
\end{equation}
where 
the definitions of \(\bm{u}\), \(\bm{X}\), \(\bm{X}'\), \(\bm{a}\), \(\bm{a}'\) and \(\bm{r}'\) are shown below
:
\begin{equation}
	\bm{u}=u(\bm{x})=
	\begin{bmatrix}
		u(x_1) & u(x_2) & \cdots & u(x_{m(I)})
	\end{bmatrix}^T,
\end{equation}
\begin{equation}
	\begin{split}
		\bm{X} &=
		\begin{bmatrix}
			\psi_1(\bm{x}) & \psi_2(\bm{x}) & \cdots & \psi_{N_I}(\bm{x})
		\end{bmatrix} \\
		&=
		\begin{bmatrix}
			\psi_1(x_1)      & \psi_2(x_1)      & \cdots & \psi_{N_I}(x_1) \\
			\psi_1(x_2)      & \psi_2(x_2)      & \cdots & \psi_{N_I}(x_2) \\
			\vdots           & \vdots           & \ddots & \vdots      \\
			\psi_1(x_{m(I)}) & \psi_2(x_{m(I)}) & \cdots & \psi_{N_I}(x_{m(I)})
		\end{bmatrix},
	\end{split}
\end{equation}
\begin{equation}
	\bm{X}' =
	\begin{bmatrix}
		\psi_1(\bm{x}) & \psi_2(\bm{x}) & \cdots & \psi_{N'}(\bm{x})
	\end{bmatrix},
\end{equation}
\begin{equation}
	\bm{a} =
	\begin{bmatrix}
		a_1 & a_2 & \cdots & a_{N_I}
	\end{bmatrix}^T,
\end{equation}
\begin{equation}
	\bm{a}' =
	\begin{bmatrix}
		a_1 & a_2 & \cdots & a_{N'}
	\end{bmatrix}^T,
\end{equation}
\begin{equation}
	\bm{r}' =
	\begin{bmatrix}
		\psi_{N'+1}(\bm{x}) & \psi_{N'+2}(\bm{x}) & \cdots & \psi_{N_I}(\bm{x})
	\end{bmatrix}
	\begin{bmatrix}
		a_{N'+1} & a_{N'+2} & \cdots & a_{N_I}
	\end{bmatrix}^T.
\end{equation}
Here $(\cdot)'$ is connected to truncating the POD space with $N_I$ modes to the leading $N'$ modes. Before moving on to GPOD reconstruction, we denote
\begin{equation}
	\bar{\bm{u}} = u(\bar{\bm{x}}), \quad \tilde{\bm{u}} = u(\tilde{\bm{x}}),
\end{equation}
and
\begin{equation}
	\begin{split}
		\bar{\bm{X}} &=
		\begin{bmatrix}
			\psi_1(\bar{\bm{x}}) & \psi_2(\bar{\bm{x}}) & \cdots & \psi_{N_I}(\bar{\bm{x}})
		\end{bmatrix} \\
		&=
		\begin{bmatrix}
			\psi_1(x_{i_1})     & \psi_2(x_{i_1})      & \cdots & \psi_{N_I}(x_{i_1}) \\
			\psi_1(x_{i_2})     & \psi_2(x_{i_2})      & \cdots & \psi_{N_I}(x_{i_2}) \\
			\vdots              & \vdots               & \ddots & \vdots          \\
			\psi_1(x_{i_{m(G)}}) & \psi_2(x_{i_{m(G)}}) & \cdots & \psi_{N_I}(x_{i_{m(G)}})
		\end{bmatrix}, \\
		\tilde{\bm{X}} &=
		\begin{bmatrix}
			\psi_1(\tilde{\bm{x}}) & \psi_2(\tilde{\bm{x}}) & \cdots & \psi_{N_I}(\tilde{\bm{x}})
		\end{bmatrix} \\
		&=
		\begin{bmatrix}
			\psi_1(x_{j_1})      & \psi_2(x_{j_1})      & \cdots & \psi_{N_I}(x_{j_1}) \\
			\psi_1(x_{j_2})      & \psi_2(x_{j_2})      & \cdots & \psi_{N_I}(x_{j_2}) \\
			\vdots               & \vdots               & \ddots & \vdots          \\
			\psi_1(x_{j_{m(S)}}) & \psi_2(x_{j_{m(S)}}) & \cdots & \psi_{N_I}(x_{j_{m(S)}})
		\end{bmatrix}.
	\end{split}
\end{equation}
Besides, \(\bar{\bm{X}}'\), \(\tilde{\bm{X}}'\), \(\bar{\bm{r}}'\) and \(\tilde{\bm{r}}'\) can be similarly defined.

\begin{figure}
	\centering
	\includegraphics[width=0.7\linewidth]{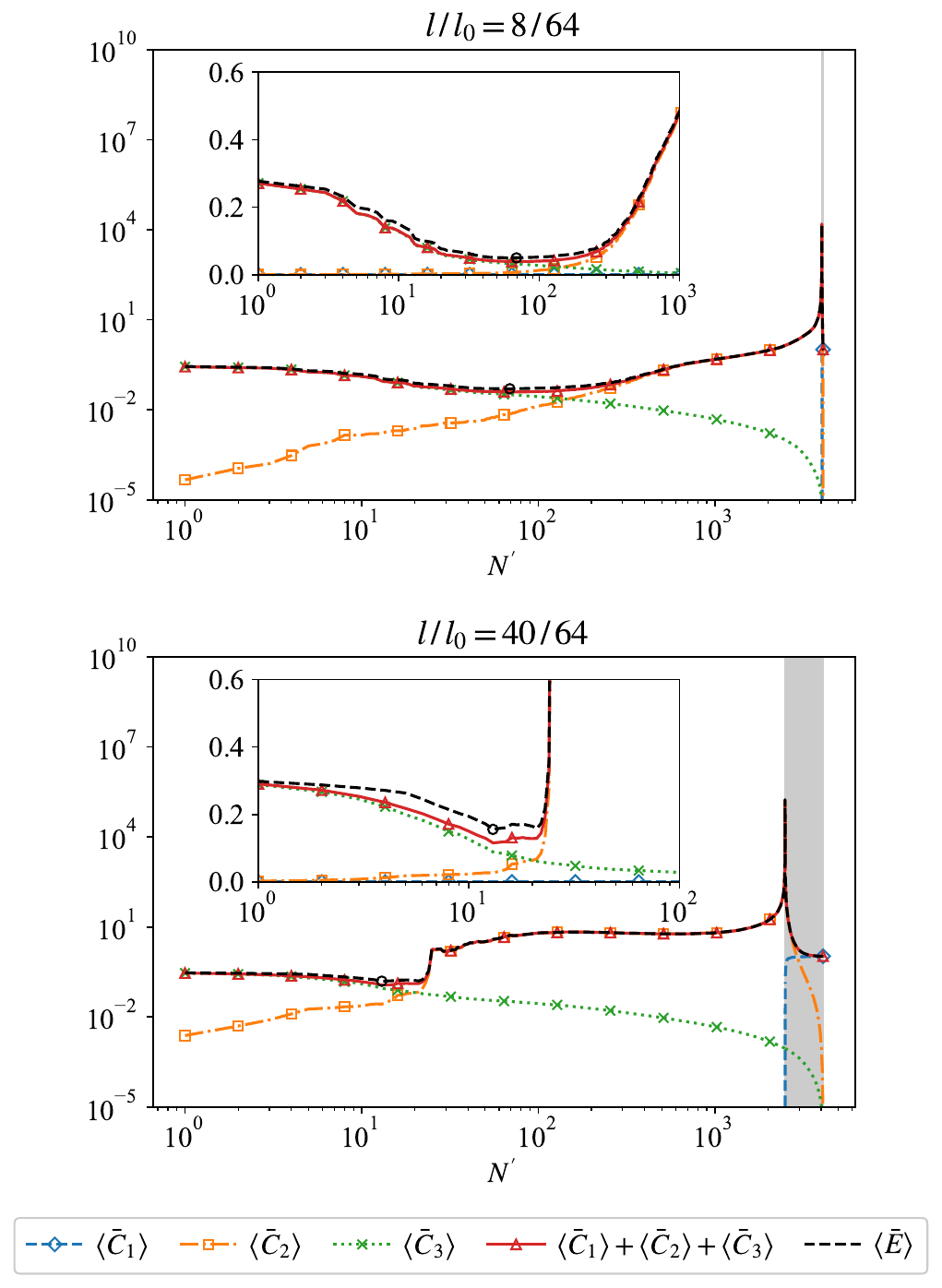}
	\caption{GPOD reconstruction error with its different contributions as functions of the number of POD modes, for a square gap with different gap sizes. The corresponding plots in the lin-log scale are shown in the insets. The black circles indicate the optimal $N'$ with the smallest reconstruction error. The range of $N'$ where the arbitrariness of $\bm{w}'$ takes effect is indicated by the gray area.}
	\label{fig:Error_analysis-sq_gap40}
\end{figure}

To conduct GPOD reconstruction, we minimize the error in the measurement region $S$ given by (\ref{equ:err_gpod}),
\begin{equation}
	\tilde{E}=\int_S|u(\bm{x})-\sum_{n=1}^{N'}a_{n}^{(p)}\psi_n(\bm{x})|^2\,\mathrm{d}\bm{x}=\|\tilde{\bm{u}}-\tilde{\bm{X}}'\bm{a}'^{(p)}\|^2,
\end{equation}
and the best fit of coefficients is given as \citep{penrose1956best, planitz19793}
\begin{equation} \label{equ:coef_b}
	\bm{a}'^{(p)} = \tilde{\bm{X}}'_+\tilde{\bm{u}} + (\bm{I}' - \tilde{\bm{X}}'_+\tilde{\bm{X}}')\bm{w}',
\end{equation}
where \(\bm{I}' \in \mathbb{R}^{N'\times N'}\) is an identity matrix, \(\tilde{\bm{X}}'_+\) is the pseudoinverse of \(\tilde{\bm{X}}'\) satisfying the Moore-Penrose conditions, and \(\bm{w}' \in \mathbb{R}^{N'\times1}\) is an arbitrary vector. Then the reconstructed field in $G$ is obtained from
\begin{equation} \label{equ:rec_field}
	\bar{\bm{u}}^{(p)} = \bar{\bm{X}}'\bm{a}'^{(p)},
\end{equation}
and the reconstruction error is
\begin{equation} \label{equ:rec_err1}
	\bar{E} = \| \bar{\bm{u}} - \bar{\bm{u}}^{(p)} \|^2 = \| \bar{\bm{X}}'[ (\bm{I}' - \tilde{\bm{X}}'_+\tilde{\bm{X}}')(\bm{a}' - \bm{w}') - \tilde{\bm{X}}'_+\tilde{\bm{r}}' ] + \bar{\bm{r}}' \|^2 = \|\bar{\bm{e}}_1 + \bar{\bm{e}}_2 + \bar{\bm{e}}_3\|^2,
\end{equation}
where
\begin{equation} \label{equ:rec_err2}
	\bar{\bm{e}}_1 = \bar{\bm{X}}'(\bm{I}' - \tilde{\bm{X}}'_+\tilde{\bm{X}}')(\bm{a}' - \bm{w}'), \quad 
	\bar{\bm{e}}_2 = -\bar{\bm{X}}'\tilde{\bm{X}}'_+\tilde{\bm{r}}', \quad
	\bar{\bm{e}}_3 = \bar{\bm{r}}'.
\end{equation}
Equations (\ref{equ:rec_err1}) and (\ref{equ:rec_err2}) show that the reconstruction error depends on three terms, the contributions of which can be calculated as
\begin{equation}
	\bar{C}_1 = \|\bar{\bm{e}}_1\|^2, \quad
	\bar{C}_2 = \|\bar{\bm{e}}_2\|^2, \quad
	\bar{C}_3 = \|\bar{\bm{e}}_3\|^2.
\end{equation}
For a square gap with sizes $l/l_0=8/64$ and $40/64$, figure \ref{fig:Error_analysis-sq_gap40} shows \(\langle \bar{C}_1 \rangle\), \(\langle \bar{C}_2 \rangle\), \(\langle \bar{C}_3 \rangle\) and \(\langle \bar{E} \rangle\) as functions of \(N'\), where the angle brackets represent the average over training data. Quantities are normalized by $A_GE_{u_G}$.
It shows that $\langle \bar{C}_1 \rangle$ is always zero when $N'$ is smaller than a threshold, $N'_\mathrm{c}$, because 
in this case $\tilde{\bm{X}}'$ is invertible (with the condition number less than $10^{10}$) and thus $\bm{I}' - \tilde{\bm{X}}'_+\tilde{\bm{X}}' = \bm{0}$ in (\ref{equ:rec_err2}). The arbitrariness of $\bm{w}'$ only takes effect when $N'$ is larger than 
$N'_\mathrm{c}$, in which case $\tilde{\bm{X}}'$ is not invertible and $\langle \bar{C}_1 \rangle$ is not zero. In figure \ref{fig:Error_analysis-sq_gap40} we use the gray area to indicate this range of $N'$ and plot $\langle \bar{C}_1 \rangle$ and $\langle \tilde{E} \rangle$ with \(\bm{w}' = \bm{0}\). When $N'$ increases from zero, $\langle \bar{C}_3 \rangle$ always decreases as it represents the truncation error of POD expansion, while $\langle \bar{C}_2 \rangle$ increases at $N'<N'_\mathrm{c}$ and decreases at $N'>N'_\mathrm{c}$. 
Because of the trade-off between different error components, there exists an optimal $N'$ with the smallest reconstruction error $\langle\bar{E}\rangle$, which will be used in the testing process.

\section{GPOD reconstruction with Lasso regularization}\label{appB}
Different from using dimension reduction (DR) to keep only the leading POD modes, GPOD can use the complete POD decomposition for reconstruction 
\begin{equation}
	u_G^{(p)}(\bm{x})=\sum_{n=1}^{N_I}a_{n}^{(p)}\psi_n(\bm{x})\qquad(\bm{x}\in G),
\end{equation}
and minimize the distance between the measurements and the POD decomposition with the help of Lasso regularization \citep{tibshirani1996regression}:
\begin{equation}
	\tilde{E}_{L_1} = \int_S | u_S(\bm{x}) - \sum_{n=1}^{N_I} a_{n}^{(p)}\psi_n(\bm{x}) |^2 \, \mathrm{d}\bm{x} + \alpha \sum_{n=1}^{N_I} |a_{n}^{(p)}|.
\end{equation}
Lasso penalizes the $L_1$ norm of the coefficients and tends to produce some coefficients that are exactly zero, which is similar to finding a best subset of POD modes that does not necessarily consist of the leading ones. The hyper-parameter $\alpha$ controls regularization strength and we estimate $\alpha$ by five-fold cross-validation \citep{efron1994introduction} with the data in $S$ during the reconstruction process.

\begin{figure}
	\centering
	\includegraphics[width=1.0\linewidth]{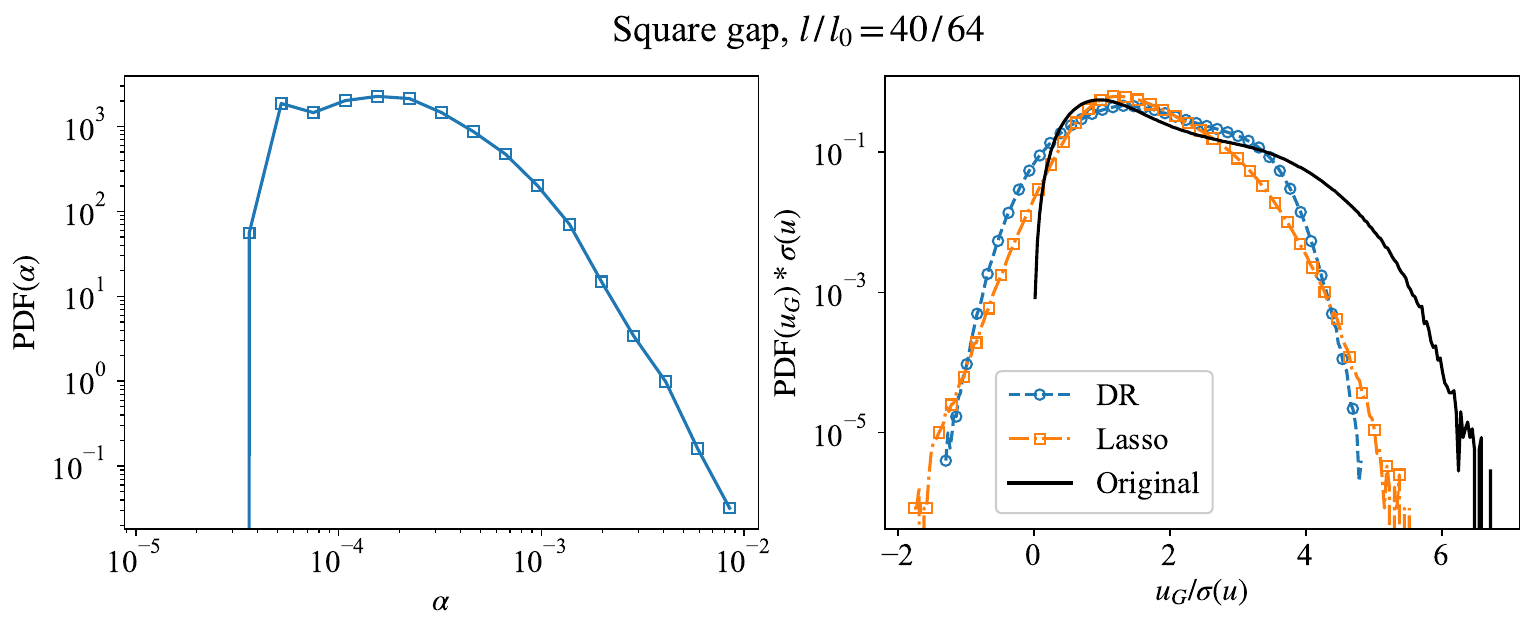}
	\caption{PDF of the estimated value of $\alpha$ over the test data for Lasso regression (left) and PDFs of the velocity module from the ground truth and that from the missing region obtained from GPOD with DR and Lasso (right) for a square gap of size $l/l_0=40/64$.}
	\label{fig:alpha-PDF_GPOD_DR_Lasso-sq_gap40}
\end{figure}

\begin{figure}
	\centering
	\includegraphics[width=1.0\linewidth]{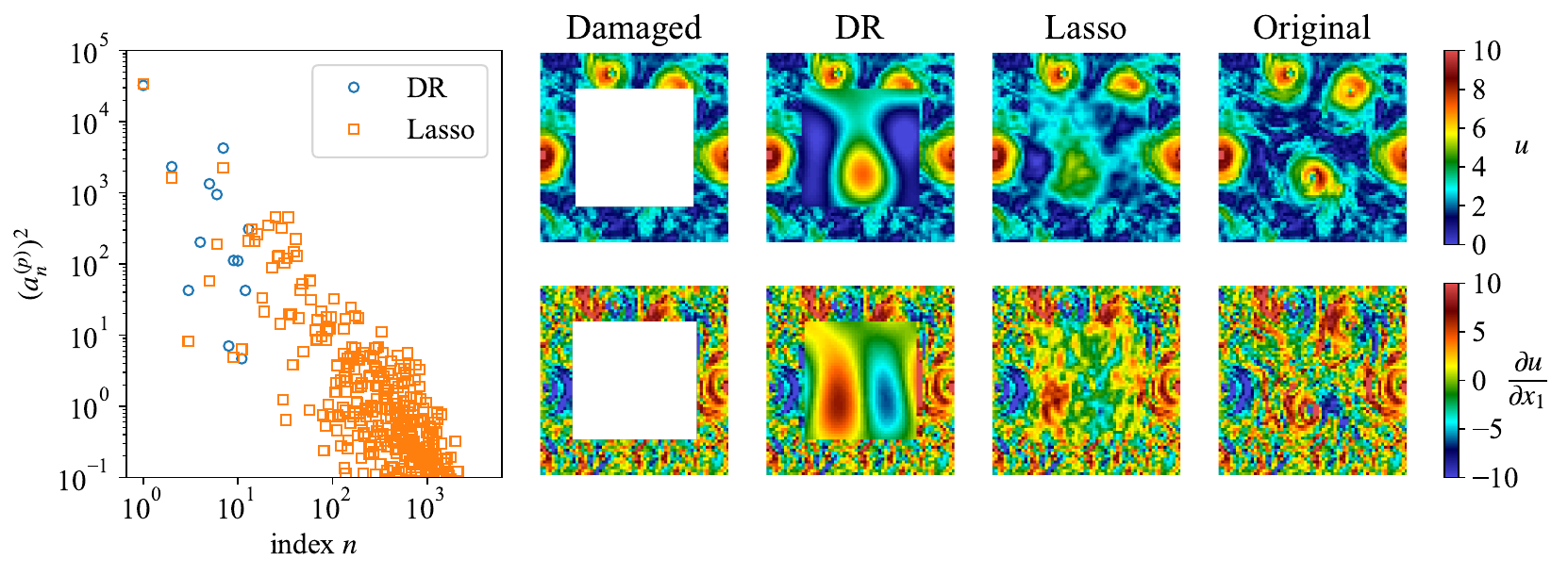}
	\caption{The spectra of the predicted POD coefficients obtained from the GPOD with DR and Lasso for an instantaneous field with a square gap of size $l/l_0=40/64$ (left). The corresponding damaged, reconstructed and original velocity module fields with their gradient fields are shown on the right.}
	\label{fig:Reconstruction-GPOD_DR_Lasso-sq_gap40}
\end{figure}

With this approach, we conducted a reconstruction experiment for a square gap of size $l/l_0=40/64$ for illustration 
and it is not our intention to perform a systematic investigation at changing the geometry and area of the gap.
Figure \ref{fig:alpha-PDF_GPOD_DR_Lasso-sq_gap40} (left) shows the PDF of the estimated value of $\alpha$ over the test data for Lasso regression.
Table \ref{tab:MSE-JSD-GPOD_DR_Lasso-sq_gap40} shows that the GPOD reconstructions with DR and Lasso give similar values of $\mathrm{MSE}(u_G)$ and $\mathrm{JSD}(u_G)$.
\begin{table}
	\begin{center}
		\def~{\hphantom{0}}
		\begin{tabular}{ccc}
			& $\mathrm{MSE}(u_G)$ & $\mathrm{JSD}(u_G)$\\[5pt]
			DR & $0.17_{-0.02}^{+0.01}$ & $0.048_{-0.001}^{+0.001}$\\[5pt]
			Lasso & $0.20_{-0.02}^{+0.02}$ & $0.049_{-0.001}^{+0.001}$\\
		\end{tabular}
		\caption{The MSE and the JS divergence between PDFs for the original and generated velocity module inside the missing region, obtained from GPOD with DR and Lasso for a square gap of size \(l/l_0=40/64\). The mean value and the error bound are calculated over test batches of size 128 for MSE and 2048 for JS divergence.}
		\label{tab:MSE-JSD-GPOD_DR_Lasso-sq_gap40}
	\end{center}
\end{table}
Figure \ref{fig:alpha-PDF_GPOD_DR_Lasso-sq_gap40} (right) also shows that the PDFs of their predicted velocity module are comparable. 
The difference between DR and Lasso can be illustrated by the spectra of the predicted POD coefficients of an instantaneous field with a square gap of size $l/l_0=40/64$, as shown in figure \ref{fig:Reconstruction-GPOD_DR_Lasso-sq_gap40} (left). DR gives a nonzero spectrum up to $N'=12$, while Lasso selects both large- and small-scale modes with a wide range of indices. This can be further shown with the reconstruction in figure \ref{fig:Reconstruction-GPOD_DR_Lasso-sq_gap40} (right), where DR only predicts `smooth' structures given by the leading POD modes and Lasso generates predictions with multiple scales.

\section{}\label{appC}
This appendix contains details of 
the GAN used in this study, of which the architecture is shown in figure \ref{fig:GAN_schematic_diagram}. For square gap with different sizes $l=8$, $16$, $24$, $32$, $40$, $50$, $60$ and $62$, we use different kernel sizes of the last layer of generator and the first layer of discriminator, $k=8$, $4$, $18$, $2$, $25$, $15$, $5$ and $3$. This can be obtained from the relation for the corresponding unpadded convolution (up-convolution) layer, $l=(64-k)/s+1$, as both $k$ and the stride $s$ are integers. For random gappiness, we use $l=64$ and $k=1$ but the $L_2$ loss is only computed in the gap. Moreover, the whole output of generator is used as the input of discriminator. To generate positive output (velocity module), ReLU is adopted as the activation function for the last layer of generator. The negative slope of the leaky ReLU activation function is empirically chosen as 0.2 for other convolution (up-convolution) layers.
As illustrated in \S\ref{sec:depen}, we can pick the adversarial ratio to obtain a good compromise between MSE and the reconstructed turbulent statistics, which gives $\lambda_{adv}=10^{-2}$ for a central square gap and $\lambda_{adv}=10^{-3}$ for random gappiness.

\begin{figure}
	\centering
	\includegraphics[width=1.0\linewidth]{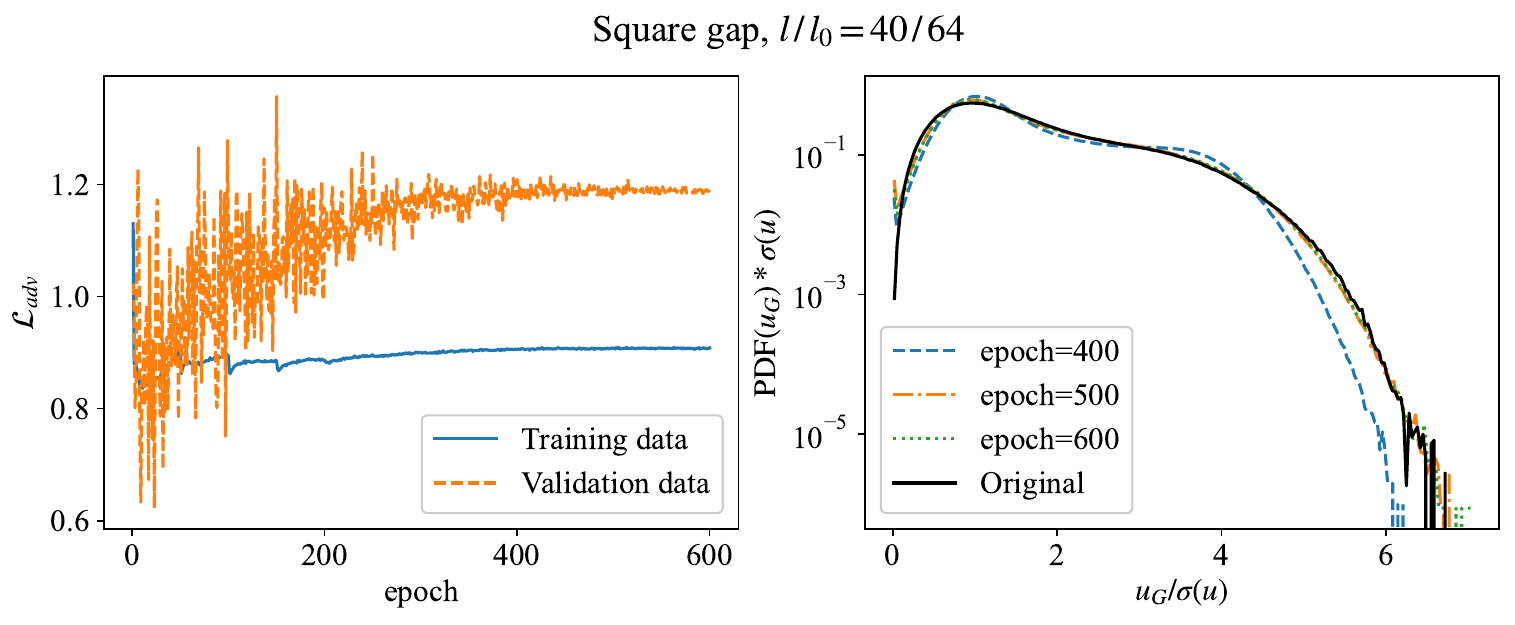}
	\caption{The adversarial loss as a function of epoch (left) and PDFs of the predictions in the missing region at different epochs and the one of the ground truth over the whole region for the validation data (right). Results are obtained from the training process of the GAN for a square gap of size $l/l_0=40/64$.}
	\label{fig:Training_process}
\end{figure}

We train the generator and discriminator together with Adam optimizer \citep{kingma2014adam}, where the learning rate of generator is twice that of discriminator. To improve the stability of training, a staircase-decay schedule is adopted to the learning rate. It decays with a rate of 0.5 every 50 epochs for 11 times, corresponding to the maximum epoch equal to 600. We choose a batch size of 128 and the initial learning rate of generator as $10^{-3}$. Figure \ref{fig:Training_process} shows the training process of the GAN for a $l/l_0=40/64$ central square gap. As training proceeds, the adversarial loss saturates at fixed values (figure \ref{fig:Training_process}(a)), while the predicted PDF gets closer to the ground truth for the validation data (figure \ref{fig:Training_process}(b)). This indicates the training convergence.

\section{}\label{appD}
To simplify the notation, we denote $x$ and $y$ respectively as the predicted and original fields inside the missing region. Here we use the angle brackets as the average over the gap and over the test data,
\begin{equation}
	\langle \cdot \rangle = \frac{1}{N_\mathrm{test}}\sum_{c=1}^{N_\mathrm{test}}(\frac{1}{A_G} \int_G (\cdot)^c \, \mathrm{d}\bm{x}),
\end{equation}
where $c$ is the index of a flow frame in the test data.
As shown in \S\ref{subsec:large}, the baseline MSE comes from uncorrelated predictions that are statistically consistent with the ground truth. Therefore, we have $\langle xy\rangle=\langle x\rangle\langle y\rangle$ because of the uncorrelation and the statistical consistency gives that $\langle x\rangle=\langle y\rangle$ and $\langle x^2\rangle=\langle y^2\rangle$. With the simplified notation, (\ref{equ:MSE}) can be rewritten as
\begin{equation}
	\mathrm{MSE}=\frac{\langle(x-y)^2\rangle}{\sqrt{\langle x^2\rangle}\sqrt{\langle y^2\rangle}}=\frac{\langle x^2\rangle-2\langle xy\rangle+\langle y^2\rangle}{\sqrt{\langle x^2\rangle}\sqrt{\langle y^2\rangle}}.
\end{equation}
Then using the relations above we can obtain the baseline MSE,
\begin{equation}
	\mathrm{MSE}_b=\frac{2(\langle x^2\rangle-\langle x\rangle^2)}{\langle x^2\rangle}=2\left(1-\frac{\langle x\rangle^2}{\langle x^2\rangle}\right).
\end{equation}
For the velocity module, with its mean value $\langle u_G\rangle$ and the mean energy $\langle u_G^2\rangle$ in the gap, we have the estimate $\mathrm{MSE}_b(u_G)\approx0.5358$. For the gradient, as $\langle\p u_G/\p x_1\rangle=0$ resulted from the periodicity, one can obtain $\mathrm{MSE}_b(\p u_G/\p x_1)\approx2$.

\bibliographystyle{jfm}
\bibliography{jfm}

\begin{thebibliography}{82}
\expandafter\ifx\csname natexlab\endcsname\relax\def\natexlab#1{#1}\fi
\def\au#1{#1} \def\ed#1{#1} \def\yr#1{#1}\def\at#1{#1}\def\jt#1{\textit{#1}}
  \def\bt#1{#1}\def\bvol#1{\textbf{#1}} \def\vol#1{#1} \def\pg#1{#1}
  \def\publ#1{#1}\def\arxiv#1{#1}\def\org#1{#1}\def\st#1{\textit{#1}}

\bibitem[Alexakis \& Biferale(2018)]{alexakis2018cascades}
{\sc \au{Alexakis, Alexandros} \& \au{Biferale, Luca}} \yr{2018}  \at{Cascades
  and transitions in turbulent flows}.  \jt{Physics Reports}  \bvol{767},
  \pg{1--101}.

\bibitem[Asch {\em et~al.\/}(2016)Asch, Bocquet \& Nodet]{asch2016data}
{\sc \au{Asch, Mark}, \au{Bocquet, Marc} \& \au{Nodet, Ma{\"e}lle}} \yr{2016}
  {\em Data assimilation: methods, algorithms, and applications\/}.
  \publ{SIAM}.

\bibitem[Baral {\em et~al.\/}(2018)Baral, Fuentes \& Kreinovich]{baral2018deep}
{\sc \au{Baral, Chitta}, \au{Fuentes, Olac} \& \au{Kreinovich, Vladik}}
  \yr{2018}  \at{Why deep neural networks: a possible theoretical explanation}.
   \bt{In {\em Constraint programming and decision making: theory and
  applications\/}},  \pg{pp. 1--5}.  \publ{Springer}.

\bibitem[Bell {\em et~al.\/}(2009)Bell, Lefebvre, Le~Traon, Smith \&
  Wilmer-Becker]{bell2009godae}
{\sc \au{Bell, MichAEl~J}, \au{Lefebvre, Michel}, \au{Le~Traon, Pierre-Yves},
  \au{Smith, Neville} \& \au{Wilmer-Becker, Kirsten}} \yr{2009}  \at{Godae: the
  global ocean data assimilation experiment}.  \jt{Oceanography}
  \bvol{22}~(3),  \pg{14--21}.

\bibitem[Biferale {\em et~al.\/}(2020)Biferale, Bonaccorso, Buzzicotti \&
  di~Leoni]{biferale2020turb}
{\sc \au{Biferale, Luca}, \au{Bonaccorso, Fabio}, \au{Buzzicotti, Michele} \&
  \au{di~Leoni, P~Clark}} \yr{2020}  \at{Turb-rot. a large database of 3d and
  2d snapshots from turbulent rotating flows}.  \jt{arXiv preprint
  arXiv:2006.07469} .

\bibitem[Bor{\'e}e(2003)]{boree2003extended}
{\sc \au{Bor{\'e}e, J}} \yr{2003}  \at{Extended proper orthogonal
  decomposition: a tool to analyse correlated events in turbulent flows}.
  \jt{Experiments in fluids}  \bvol{35}~(2),  \pg{188--192}.

\bibitem[Brunton \& Noack(2015)]{brunton2015closed}
{\sc \au{Brunton, Steven~L} \& \au{Noack, Bernd~R}} \yr{2015}  \at{Closed-loop
  turbulence control: Progress and challenges}.  \jt{Applied Mechanics Reviews}
   \bvol{67}~(5).

\bibitem[Buzzicotti {\em et~al.\/}(2016)Buzzicotti, Bhatnagar, Biferale,
  Lanotte \& Ray]{buzzicotti2016lagrangian}
{\sc \au{Buzzicotti, Michele}, \au{Bhatnagar, Akshay}, \au{Biferale, Luca},
  \au{Lanotte, Alessandra~S} \& \au{Ray, Samriddhi~Sankar}} \yr{2016}
  \at{Lagrangian statistics for navier--stokes turbulence under fourier-mode
  reduction: fractal and homogeneous decimations}.  \jt{New Journal of Physics}
   \bvol{18}~(11),  \pg{113047}.

\bibitem[Buzzicotti \& Bonaccorso(2022)]{buzzicotti2022inferring}
{\sc \au{Buzzicotti, Michele} \& \au{Bonaccorso, Fabio}} \yr{2022}
  \at{Inferring turbulent environments via machine learning}.  \jt{The European
  Physical Journal E}  \bvol{45}~(12),  \pg{102}.

\bibitem[Buzzicotti {\em et~al.\/}(2021)Buzzicotti, Bonaccorso, Di~Leoni \&
  Biferale]{buzzicotti2021reconstruction}
{\sc \au{Buzzicotti, Michele}, \au{Bonaccorso, Fabio}, \au{Di~Leoni, P~Clark}
  \& \au{Biferale, Luca}} \yr{2021}  \at{Reconstruction of turbulent data with
  deep generative models for semantic inpainting from turb-rot database}.
  \jt{Physical Review Fluids}  \bvol{6}~(5),  \pg{050503}.

\bibitem[Buzzicotti {\em et~al.\/}(2018)Buzzicotti, Clark Di~Leoni \&
  Biferale]{buzzicotti2018inverse}
{\sc \au{Buzzicotti, Michele}, \au{Clark Di~Leoni, Patricio} \& \au{Biferale,
  Luca}} \yr{2018}  \at{On the inverse energy transfer in rotating turbulence}.
   \jt{The European Physical Journal E}  \bvol{41}~(11),  \pg{1--8}.

\bibitem[Choi {\em et~al.\/}(1994)Choi, Moin \& Kim]{choi1994active}
{\sc \au{Choi, Haecheon}, \au{Moin, Parviz} \& \au{Kim, John}} \yr{1994}
  \at{Active turbulence control for drag reduction in wall-bounded flows}.
  \jt{Journal of Fluid Mechanics}  \bvol{262},  \pg{75--110}.

\bibitem[Clark Di~Leoni {\em et~al.\/}(2022)Clark Di~Leoni, Agarwal, Zaki,
  Meneveau \& Katz]{di2022reconstructing}
{\sc \au{Clark Di~Leoni, Patricio}, \au{Agarwal, Karuna}, \au{Zaki, Tamer},
  \au{Meneveau, Charles} \& \au{Katz, Joseph}} \yr{2022}  \at{Reconstructing
  velocity and pressure from sparse noisy particle tracks using
  physics-informed neural networks}.  \jt{arXiv preprint arXiv:2210.04849} .

\bibitem[Cohen \& Kundu(2004)]{cohen2004fluid}
{\sc \au{Cohen, Ira~M} \& \au{Kundu, Pijush~K}} \yr{2004} {\em Fluid
  mechanics\/}.  \publ{Elsevier}.

\bibitem[Dabiri \& Pecora(2020)]{dabiri2020particle}
{\sc \au{Dabiri, Dana} \& \au{Pecora, Charles}} \yr{2020} {\em Particle
  tracking velocimetry\/}, ,  \vol{vol. 785}.  \publ{IOP Publishing Bristol}.

\bibitem[Deng {\em et~al.\/}(2019)Deng, He, Liu \& Kim]{deng2019super}
{\sc \au{Deng, Zhiwen}, \au{He, Chuangxin}, \au{Liu, Yingzheng} \& \au{Kim,
  Kyung~Chun}} \yr{2019}  \at{Super-resolution reconstruction of turbulent
  velocity fields using a generative adversarial network-based artificial
  intelligence framework}.  \jt{Physics of Fluids}  \bvol{31}~(12),
  \pg{125111}.

\bibitem[Di~Leoni {\em et~al.\/}(2020)Di~Leoni, Alexakis, Biferale \&
  Buzzicotti]{di2020phase}
{\sc \au{Di~Leoni, P~Clark}, \au{Alexakis, Alexandros}, \au{Biferale, L} \&
  \au{Buzzicotti, M}} \yr{2020}  \at{Phase transitions and flux-loop metastable
  states in rotating turbulence}.  \jt{Physical Review Fluids}  \bvol{5}~(10),
  \pg{104603}.

\bibitem[Discetti {\em et~al.\/}(2019)Discetti, Bellani, {\"O}rl{\"u},
  Serpieri, Vila, Raiola, Zheng, Mascotelli, Talamelli \&
  Ianiro]{discetti2019characterization}
{\sc \au{Discetti, Stefano}, \au{Bellani, Gabriele}, \au{{\"O}rl{\"u}, Ramis},
  \au{Serpieri, Jacopo}, \au{Vila, Carlos~Sanmiguel}, \au{Raiola, Marco},
  \au{Zheng, Xiaobo}, \au{Mascotelli, Lucia}, \au{Talamelli, Alessandro} \&
  \au{Ianiro, Andrea}} \yr{2019}  \at{Characterization of very-large-scale
  motions in high-re pipe flows}.  \jt{Experimental Thermal and Fluid Science}
  \bvol{104},  \pg{1--8}.

\bibitem[Efron \& Tibshirani(1994)]{efron1994introduction}
{\sc \au{Efron, Bradley} \& \au{Tibshirani, Robert~J}} \yr{1994} {\em An
  introduction to the bootstrap\/}.  \publ{CRC press}.

\bibitem[Everson \& Sirovich(1995)]{everson1995karhunen}
{\sc \au{Everson, Richard} \& \au{Sirovich, Lawrence}} \yr{1995}
  \at{Karhunen--loeve procedure for gappy data}.  \jt{JOSA A}  \bvol{12}~(8),
  \pg{1657--1664}.

\bibitem[Fahland {\em et~al.\/}(2021)Fahland, Stroh, Frohnapfel, Atzori,
  Vinuesa, Schlatter \& Gatti]{fahland2021investigation}
{\sc \au{Fahland, Georg}, \au{Stroh, Alexander}, \au{Frohnapfel, Bettina},
  \au{Atzori, Marco}, \au{Vinuesa, Ricardo}, \au{Schlatter, Philipp} \&
  \au{Gatti, Davide}} \yr{2021}  \at{Investigation of blowing and suction for
  turbulent flow control on airfoils}.  \jt{AIAA Journal}  \bvol{59}~(11),
  \pg{4422--4436}.

\bibitem[Frisch(1995)]{frisch1995turbulence}
{\sc \au{Frisch, Uriel}} \yr{1995}  \at{Turbulence: The legacy of an
  kolmogorov} .

\bibitem[Frisch {\em et~al.\/}(2008)Frisch, Kurien, Pandit, Pauls, Ray, Wirth
  \& Zhu]{frisch2008hyperviscosity}
{\sc \au{Frisch, Uriel}, \au{Kurien, Susan}, \au{Pandit, Rahul}, \au{Pauls,
  Walter}, \au{Ray, Samriddhi~Sankar}, \au{Wirth, Achim} \& \au{Zhu,
  Jian-Zhou}} \yr{2008}  \at{Hyperviscosity, galerkin truncation, and
  bottlenecks in turbulence}.  \jt{Physical review letters}  \bvol{101}~(14),
  \pg{144501}.

\bibitem[Fukami {\em et~al.\/}(2019)Fukami, Fukagata \& Taira]{fukami2019super}
{\sc \au{Fukami, Kai}, \au{Fukagata, Koji} \& \au{Taira, Kunihiko}} \yr{2019}
  \at{Super-resolution reconstruction of turbulent flows with machine
  learning}.  \jt{Journal of Fluid Mechanics}  \bvol{870},  \pg{106--120}.

\bibitem[Fukami {\em et~al.\/}(2021)Fukami, Fukagata \&
  Taira]{fukami2021machine}
{\sc \au{Fukami, Kai}, \au{Fukagata, Koji} \& \au{Taira, Kunihiko}} \yr{2021}
  \at{Machine-learning-based spatio-temporal super resolution reconstruction of
  turbulent flows}.  \jt{Journal of Fluid Mechanics}  \bvol{909}.

\bibitem[Fukunaga(2013)]{fukunaga2013introduction}
{\sc \au{Fukunaga, Keinosuke}} \yr{2013} {\em Introduction to statistical
  pattern recognition\/}.  \publ{Elsevier}.

\bibitem[Garcia(2011)]{garcia2011fast}
{\sc \au{Garcia, Damien}} \yr{2011}  \at{A fast all-in-one method for automated
  post-processing of piv data}.  \jt{Experiments in fluids}  \bvol{50}~(5),
  \pg{1247--1259}.

\bibitem[Godeferd \& Moisy(2015)]{godeferd2015structure}
{\sc \au{Godeferd, Fabien~S} \& \au{Moisy, Fr{\'e}d{\'e}ric}} \yr{2015}
  \at{Structure and dynamics of rotating turbulence: a review of recent
  experimental and numerical results}.  \jt{Applied Mechanics Reviews}
  \bvol{67}~(3),  \pg{030802}.

\bibitem[Goodfellow {\em et~al.\/}(2014)Goodfellow, Pouget-Abadie, Mirza, Xu,
  Warde-Farley, Ozair, Courville \& Bengio]{goodfellow2014generative}
{\sc \au{Goodfellow, Ian}, \au{Pouget-Abadie, Jean}, \au{Mirza, Mehdi}, \au{Xu,
  Bing}, \au{Warde-Farley, David}, \au{Ozair, Sherjil}, \au{Courville, Aaron}
  \& \au{Bengio, Yoshua}} \yr{2014}  \at{Generative adversarial nets}.
  \jt{Advances in neural information processing systems}  \bvol{27}.

\bibitem[Guastoni {\em et~al.\/}(2021)Guastoni, G{\"u}emes, Ianiro, Discetti,
  Schlatter, Azizpour \& Vinuesa]{guastoni2021convolutional}
{\sc \au{Guastoni, Luca}, \au{G{\"u}emes, Alejandro}, \au{Ianiro, Andrea},
  \au{Discetti, Stefano}, \au{Schlatter, Philipp}, \au{Azizpour, Hossein} \&
  \au{Vinuesa, Ricardo}} \yr{2021}  \at{Convolutional-network models to predict
  wall-bounded turbulence from wall quantities}.  \jt{Journal of Fluid
  Mechanics}  \bvol{928}.

\bibitem[G{\"u}emes {\em et~al.\/}(2019)G{\"u}emes, Discetti \&
  Ianiro]{guemes2019sensing}
{\sc \au{G{\"u}emes, A}, \au{Discetti, S} \& \au{Ianiro, A}} \yr{2019}
  \at{Sensing the turbulent large-scale motions with their wall signature}.
  \jt{Physics of Fluids}  \bvol{31}~(12),  \pg{125112}.

\bibitem[G{\"u}emes {\em et~al.\/}(2021)G{\"u}emes, Discetti, Ianiro, Sirmacek,
  Azizpour \& Vinuesa]{guemes2021coarse}
{\sc \au{G{\"u}emes, Alejandro}, \au{Discetti, Stefano}, \au{Ianiro, Andrea},
  \au{Sirmacek, Beril}, \au{Azizpour, Hossein} \& \au{Vinuesa, Ricardo}}
  \yr{2021}  \at{From coarse wall measurements to turbulent velocity fields
  through deep learning}.  \jt{Physics of Fluids}  \bvol{33}~(7),  \pg{075121}.

\bibitem[Gunes \& Rist(2008)]{gunes2008use}
{\sc \au{Gunes, Hasan} \& \au{Rist, Ulrich}} \yr{2008}  \at{On the use of
  kriging for enhanced data reconstruction in a separated transitional
  flat-plate boundary layer}.  \jt{Physics of Fluids}  \bvol{20}~(10),
  \pg{104109}.

\bibitem[Gunes {\em et~al.\/}(2006)Gunes, Sirisup \&
  Karniadakis]{gunes2006gappy}
{\sc \au{Gunes, Hasan}, \au{Sirisup, Sirod} \& \au{Karniadakis, George~Em}}
  \yr{2006}  \at{Gappy data: To krig or not to krig?}  \jt{Journal of
  Computational Physics}  \bvol{212}~(1),  \pg{358--382}.

\bibitem[Gad-el Hak \& Tsai(2006)]{gad2006transition}
{\sc \au{Gad-el Hak, Mohamed} \& \au{Tsai, Her~Mann}} \yr{2006} {\em Transition
  and turbulence control\/}, ,  \vol{vol.~8}.  \publ{World Scientific}.

\bibitem[Haugen \& Brandenburg(2004)]{haugen2004inertial}
{\sc \au{Haugen, Nils Erland~L} \& \au{Brandenburg, Axel}} \yr{2004}
  \at{Inertial range scaling in numerical turbulence with hyperviscosity}.
  \jt{Physical Review E}  \bvol{70}~(2),  \pg{026405}.

\bibitem[He {\em et~al.\/}(2016)He, Zhang, Ren \& Sun]{he2016deep}
{\sc \au{He, Kaiming}, \au{Zhang, Xiangyu}, \au{Ren, Shaoqing} \& \au{Sun,
  Jian}} \yr{2016} Deep residual learning for image recognition.  \bt{In {\em
  Proceedings of the IEEE conference on computer vision and pattern
  recognition\/}},  \pg{pp. 770--778}.

\bibitem[Holmes {\em et~al.\/}(2012)Holmes, Lumley, Berkooz \&
  Rowley]{holmes2012turbulence}
{\sc \au{Holmes, Philip}, \au{Lumley, John~L}, \au{Berkooz, Gahl} \&
  \au{Rowley, Clarence~W}} \yr{2012} {\em Turbulence, coherent structures,
  dynamical systems and symmetry\/}.  \publ{Cambridge university press}.

\bibitem[Hornik(1991)]{hornik1991approximation}
{\sc \au{Hornik, Kurt}} \yr{1991}  \at{Approximation capabilities of multilayer
  feedforward networks}.  \jt{Neural networks}  \bvol{4}~(2),  \pg{251--257}.

\bibitem[Hosseini {\em et~al.\/}(2016)Hosseini, Martinuzzi \&
  Noack]{hosseini2016modal}
{\sc \au{Hosseini, Zahra}, \au{Martinuzzi, Robert~J} \& \au{Noack, Bernd~R}}
  \yr{2016}  \at{Modal energy flow analysis of a highly modulated wake behind a
  wall-mounted pyramid}.  \jt{Journal of Fluid Mechanics}  \bvol{798},
  \pg{717--750}.

\bibitem[van Kan \& Alexakis(2020)]{van2020critical}
{\sc \au{van Kan, Adrian} \& \au{Alexakis, Alexandros}} \yr{2020}  \at{Critical
  transition in fast-rotating turbulence within highly elongated domains}.
  \jt{Journal of Fluid Mechanics}  \bvol{899},  \pg{A33}.

\bibitem[Kim {\em et~al.\/}(2021)Kim, Kim, Won \& Lee]{kim2021unsupervised}
{\sc \au{Kim, Hyojin}, \au{Kim, Junhyuk}, \au{Won, Sungjin} \& \au{Lee,
  Changhoon}} \yr{2021}  \at{Unsupervised deep learning for super-resolution
  reconstruction of turbulence}.  \jt{Journal of Fluid Mechanics}  \bvol{910}.

\bibitem[Kim \& Lee(2020)]{kim2020prediction}
{\sc \au{Kim, Junhyuk} \& \au{Lee, Changhoon}} \yr{2020}  \at{Prediction of
  turbulent heat transfer using convolutional neural networks}.  \jt{Journal of
  Fluid Mechanics}  \bvol{882},  \pg{A18}.

\bibitem[Kingma \& Ba(2014)]{kingma2014adam}
{\sc \au{Kingma, Diederik~P} \& \au{Ba, Jimmy}} \yr{2014}  \at{Adam: A method
  for stochastic optimization}.  \jt{arXiv preprint arXiv:1412.6980} .

\bibitem[Kreinovich(1991)]{kreinovich1991arbitrary}
{\sc \au{Kreinovich, Vladik~Ya}} \yr{1991}  \at{Arbitrary nonlinearity is
  sufficient to represent all functions by neural networks: a theorem}.
  \jt{Neural networks}  \bvol{4}~(3),  \pg{381--383}.

\bibitem[Krysta {\em et~al.\/}(2011)Krysta, Blayo, Cosme \&
  Verron]{krysta2011consistent}
{\sc \au{Krysta, Monika}, \au{Blayo, Eric}, \au{Cosme, Emmanuel} \& \au{Verron,
  Jacques}} \yr{2011}  \at{A consistent hybrid variational-smoothing data
  assimilation method: Application to a simple shallow-water model of the
  turbulent midlatitude ocean}.  \jt{Monthly weather review}  \bvol{139}~(11),
  \pg{3333--3347}.

\bibitem[Le~Dimet \& Talagrand(1986)]{le1986variational}
{\sc \au{Le~Dimet, Fran{\c{c}}ois-Xavier} \& \au{Talagrand, Olivier}} \yr{1986}
   \at{Variational algorithms for analysis and assimilation of meteorological
  observations: theoretical aspects}.  \jt{Tellus A: Dynamic Meteorology and
  Oceanography}  \bvol{38}~(2),  \pg{97--110}.

\bibitem[Lee {\em et~al.\/}(1997)Lee, Kim, Babcock \&
  Goodman]{lee1997application}
{\sc \au{Lee, Changhoon}, \au{Kim, John}, \au{Babcock, David} \& \au{Goodman,
  Rodney}} \yr{1997}  \at{Application of neural networks to turbulence control
  for drag reduction}.  \jt{Physics of Fluids}  \bvol{9}~(6),  \pg{1740--1747}.

\bibitem[Li {\em et~al.\/}(2023)Li, Buzzicotti, Biferale \&
  Bonaccorso]{https://doi.org/10.48550/arxiv.2301.07541}
{\sc \au{Li, Tianyi}, \au{Buzzicotti, Michele}, \au{Biferale, Luca} \&
  \au{Bonaccorso, Fabio}} \yr{2023} Generative adversarial networks to infer
  velocity components in rotating turbulent flows.

\bibitem[Li {\em et~al.\/}(2021)Li, Buzzicotti, Biferale, Wan \&
  Chen]{Li2021gappy}
{\sc \au{Li, Tianyi}, \au{Buzzicotti, Michele}, \au{Biferale, Luca}, \au{Wan,
  Minping} \& \au{Chen, Shiyi}} \yr{2021}  \at{Reconstruction of turbulent data
  with gappy pod method}.  \jt{Chinese Journal of Theoretical and Applied
  Mechanics}  \bvol{53}~(10),  \pg{2703--2711}.

\bibitem[Liu {\em et~al.\/}(2020)Liu, Tang, Huang \& Lu]{liu2020deep}
{\sc \au{Liu, Bo}, \au{Tang, Jiupeng}, \au{Huang, Haibo} \& \au{Lu, Xi-Yun}}
  \yr{2020}  \at{Deep learning methods for super-resolution reconstruction of
  turbulent flows}.  \jt{Physics of Fluids}  \bvol{32}~(2),  \pg{025105}.

\bibitem[Matsuo {\em et~al.\/}(2021)Matsuo, Nakamura, Morimoto, Fukami \&
  Fukagata]{matsuo2021supervised}
{\sc \au{Matsuo, Mitsuaki}, \au{Nakamura, Taichi}, \au{Morimoto, Masaki},
  \au{Fukami, Kai} \& \au{Fukagata, Koji}} \yr{2021}  \at{Supervised
  convolutional network for three-dimensional fluid data reconstruction from
  sectional flow fields with adaptive super-resolution assistance}.  \jt{arXiv
  preprint arXiv:2103.09020} .

\bibitem[Maurel {\em et~al.\/}(2001)Maurel, Bor{\'e}e \&
  Lumley]{maurel2001extended}
{\sc \au{Maurel, S}, \au{Bor{\'e}e, J} \& \au{Lumley, JL}} \yr{2001}
  \at{Extended proper orthogonal decomposition: Application to jet/vortex
  interaction}.  \jt{Flow, Turbulence and Combustion}  \bvol{67}~(2),
  \pg{125--136}.

\bibitem[Militino {\em et~al.\/}(2019)Militino, Ugarte \&
  Montesino]{militino2019filling}
{\sc \au{Militino, Ana~F}, \au{Ugarte, MD} \& \au{Montesino, M}} \yr{2019}
  \at{Filling missing data and smoothing altered data in satellite imagery with
  a spatial functional procedure}.  \jt{Stochastic Environmental Research and
  Risk Assessment}  \bvol{33}~(10),  \pg{1737--1750}.

\bibitem[Myers(2002)]{myers2002interpolation}
{\sc \au{Myers, DE}} \yr{2002} Interpolation of spatial data: some theory for
  kriging.

\bibitem[Niu \& Suen(2012)]{niu2012novel}
{\sc \au{Niu, Xiao-Xiao} \& \au{Suen, Ching~Y}} \yr{2012}  \at{A novel hybrid
  cnn--svm classifier for recognizing handwritten digits}.  \jt{Pattern
  Recognition}  \bvol{45}~(4),  \pg{1318--1325}.

\bibitem[Oliver \& Webster(1990)]{oliver1990kriging}
{\sc \au{Oliver, Margaret~A} \& \au{Webster, Richard}} \yr{1990}  \at{Kriging:
  a method of interpolation for geographical information systems}.
  \jt{International Journal of Geographical Information System}  \bvol{4}~(3),
  \pg{313--332}.

\bibitem[Pathak {\em et~al.\/}(2016)Pathak, Krahenbuhl, Donahue, Darrell \&
  Efros]{pathak2016context}
{\sc \au{Pathak, Deepak}, \au{Krahenbuhl, Philipp}, \au{Donahue, Jeff},
  \au{Darrell, Trevor} \& \au{Efros, Alexei~A}} \yr{2016} Context encoders:
  Feature learning by inpainting.  \bt{In {\em Proceedings of the IEEE
  conference on computer vision and pattern recognition\/}},  \pg{pp.
  2536--2544}.

\bibitem[Penrose(1956)]{penrose1956best}
{\sc \au{Penrose, Roger}} \yr{1956} On best approximate solutions of linear
  matrix equations.  \bt{In {\em Mathematical Proceedings of the Cambridge
  Philosophical Society\/}}, ,  \vol{vol.~52},  \pg{pp. 17--19}. Cambridge
  University Press.

\bibitem[Planitz(1979)]{planitz19793}
{\sc \au{Planitz, M}} \yr{1979}  \at{3. inconsistent systems of linear
  equations}.  \jt{The Mathematical Gazette}  \bvol{63}~(425),  \pg{181--185}.

\bibitem[Pouquet {\em et~al.\/}(2018)Pouquet, Rosenberg, Marino \&
  Herbert]{pouquet2018scaling}
{\sc \au{Pouquet, Annick}, \au{Rosenberg, Duane}, \au{Marino, Raffaele} \&
  \au{Herbert, Corentin}} \yr{2018}  \at{Scaling laws for mixing and
  dissipation in unforced rotating stratified turbulence}.  \jt{Journal of
  Fluid Mechanics}  \bvol{844},  \pg{519--545}.

\bibitem[Romain {\em et~al.\/}(2014)Romain, Chatellier \&
  David]{romain2014bayesian}
{\sc \au{Romain, Leroux}, \au{Chatellier, Ludovic} \& \au{David, Laurent}}
  \yr{2014}  \at{Bayesian inference applied to spatio-temporal reconstruction
  of flows around a naca0012 airfoil}.  \jt{Experiments in fluids}
  \bvol{55}~(4),  \pg{1--19}.

\bibitem[Russakovsky {\em et~al.\/}(2015)Russakovsky, Deng, Su, Krause,
  Satheesh, Ma, Huang, Karpathy, Khosla, Bernstein {\em
  et~al.\/}]{russakovsky2015imagenet}
{\sc \au{Russakovsky, Olga}, \au{Deng, Jia}, \au{Su, Hao}, \au{Krause,
  Jonathan}, \au{Satheesh, Sanjeev}, \au{Ma, Sean}, \au{Huang, Zhiheng},
  \au{Karpathy, Andrej}, \au{Khosla, Aditya}, \au{Bernstein, Michael} \&
  \au{others}} \yr{2015}  \at{Imagenet large scale visual recognition
  challenge}.  \jt{International journal of computer vision}  \bvol{115}~(3),
  \pg{211--252}.

\bibitem[Sawford(1991)]{sawford1991reynolds}
{\sc \au{Sawford, BL}} \yr{1991}  \at{Reynolds number effects in lagrangian
  stochastic models of turbulent dispersion}.  \jt{Physics of Fluids A: Fluid
  Dynamics}  \bvol{3}~(6),  \pg{1577--1586}.

\bibitem[Seshasayanan \& Alexakis(2018)]{seshasayanan2018condensates}
{\sc \au{Seshasayanan, Kannabiran} \& \au{Alexakis, Alexandros}} \yr{2018}
  \at{Condensates in rotating turbulent flows}.  \jt{Journal of Fluid
  Mechanics}  \bvol{841},  \pg{434--462}.

\bibitem[Shen {\em et~al.\/}(2015)Shen, Li, Cheng, Zeng, Yang, Li \&
  Zhang]{shen2015missing}
{\sc \au{Shen, Huanfeng}, \au{Li, Xinghua}, \au{Cheng, Qing}, \au{Zeng, Chao},
  \au{Yang, Gang}, \au{Li, Huifang} \& \au{Zhang, Liangpei}} \yr{2015}
  \at{Missing information reconstruction of remote sensing data: A technical
  review}.  \jt{IEEE Geoscience and Remote Sensing Magazine}  \bvol{3}~(3),
  \pg{61--85}.

\bibitem[Singh {\em et~al.\/}(2001)Singh, Myatt, Addington, Banda \&
  Hall]{singh2001optimal}
{\sc \au{Singh, Sahjendra~N}, \au{Myatt, James~H}, \au{Addington, Gregory~A},
  \au{Banda, Siva} \& \au{Hall, James~K}} \yr{2001}  \at{Optimal feedback
  control of vortex shedding using proper orthogonal decomposition models}.
  \jt{J. Fluids Eng.}  \bvol{123}~(3),  \pg{612--618}.

\bibitem[Sirovich \& Kirby(1987)]{sirovich1987low}
{\sc \au{Sirovich, Lawrence} \& \au{Kirby, Michael}} \yr{1987}
  \at{Low-dimensional procedure for the characterization of human faces}.
  \jt{Josa a}  \bvol{4}~(3),  \pg{519--524}.

\bibitem[Storer {\em et~al.\/}(2022)Storer, Buzzicotti, Khatri, Griffies \&
  Aluie]{storer2022global}
{\sc \au{Storer, Benjamin~A}, \au{Buzzicotti, Michele}, \au{Khatri, Hemant},
  \au{Griffies, Stephen~M} \& \au{Aluie, Hussein}} \yr{2022}  \at{Global energy
  spectrum of the general oceanic circulation}.  \jt{Nature communications}
  \bvol{13}~(1),  \pg{5314}.

\bibitem[Subramaniam {\em et~al.\/}(2020)Subramaniam, Wong, Borker, Nimmagadda
  \& Lele]{subramaniam2020turbulence}
{\sc \au{Subramaniam, Akshay}, \au{Wong, Man~Long}, \au{Borker, Raunak~D},
  \au{Nimmagadda, Sravya} \& \au{Lele, Sanjiva~K}} \yr{2020}  \at{Turbulence
  enrichment using physics-informed generative adversarial networks}.
  \jt{arXiv preprint arXiv:2003.01907} .

\bibitem[Suzuki(2014)]{suzuki2014pod}
{\sc \au{Suzuki, Takao}} \yr{2014}  \at{Pod-based reduced-order hybrid
  simulation using the data-driven transfer function with time-resolved ptv
  feedback}.  \jt{Experiments in fluids}  \bvol{55}~(8),  \pg{1--17}.

\bibitem[Tibshirani(1996)]{tibshirani1996regression}
{\sc \au{Tibshirani, Robert}} \yr{1996}  \at{Regression shrinkage and selection
  via the lasso}.  \jt{Journal of the Royal Statistical Society: Series B
  (Methodological)}  \bvol{58}~(1),  \pg{267--288}.

\bibitem[Tinney {\em et~al.\/}(2008)Tinney, Ukeiley \& Glauser]{tinney2008low}
{\sc \au{Tinney, CE}, \au{Ukeiley, LS} \& \au{Glauser, Mark~N}} \yr{2008}
  \at{Low-dimensional characteristics of a transonic jet. part 2. estimate and
  far-field prediction}.  \jt{Journal of Fluid Mechanics}  \bvol{615},
  \pg{53--92}.

\bibitem[Torn \& Hakim(2009)]{torn2009ensemble}
{\sc \au{Torn, Ryan~D} \& \au{Hakim, Gregory~J}} \yr{2009}  \at{Ensemble data
  assimilation applied to rainex observations of hurricane katrina (2005)}.
  \jt{Monthly weather review}  \bvol{137}~(9),  \pg{2817--2829}.

\bibitem[Venturi \& Karniadakis(2004)]{venturi2004gappy}
{\sc \au{Venturi, Daniele} \& \au{Karniadakis, George~Em}} \yr{2004}  \at{Gappy
  data and reconstruction procedures for flow past a cylinder}.  \jt{Journal of
  Fluid Mechanics}  \bvol{519},  \pg{315--336}.

\bibitem[Wang {\em et~al.\/}(2016)Wang, Gao, Wang, Wei, Li \&
  Wang]{wang2016divergence}
{\sc \au{Wang, ChengYue}, \au{Gao, Qi}, \au{Wang, HongPing}, \au{Wei, RunJie},
  \au{Li, Tian} \& \au{Wang, JinJun}} \yr{2016}  \at{Divergence-free smoothing
  for volumetric piv data}.  \jt{Experiments in Fluids}  \bvol{57}~(1),
  \pg{15}.

\bibitem[Wang {\em et~al.\/}(2004)Wang, Bovik, Sheikh \&
  Simoncelli]{wang2004image}
{\sc \au{Wang, Zhou}, \au{Bovik, Alan~C}, \au{Sheikh, Hamid~R} \&
  \au{Simoncelli, Eero~P}} \yr{2004}  \at{Image quality assessment: from error
  visibility to structural similarity}.  \jt{IEEE transactions on image
  processing}  \bvol{13}~(4),  \pg{600--612}.

\bibitem[Wang \& Simoncelli(2005)]{wang2005translation}
{\sc \au{Wang, Zhou} \& \au{Simoncelli, Eero~P}} \yr{2005} Translation
  insensitive image similarity in complex wavelet domain.  \bt{In {\em
  Proceedings.(ICASSP'05). IEEE International Conference on Acoustics, Speech,
  and Signal Processing, 2005.\/}}, ,  \vol{vol.~2},  \pg{pp. ii--573}. IEEE.

\bibitem[Wen {\em et~al.\/}(2019)Wen, Li, Peng, Zhou \& Liu]{wen2019missing}
{\sc \au{Wen, Xin}, \au{Li, Ziyan}, \au{Peng, Di}, \au{Zhou, Wenwu} \& \au{Liu,
  Yingzheng}} \yr{2019}  \at{Missing data recovery using data fusion of
  incomplete complementary data sets: A particle image velocimetry
  application}.  \jt{Physics of Fluids}  \bvol{31}~(2),  \pg{025105}.

\bibitem[Yokoyama \& Takaoka(2021)]{yokoyama2021energy}
{\sc \au{Yokoyama, Naoto} \& \au{Takaoka, Masanori}} \yr{2021}  \at{Energy-flux
  vector in anisotropic turbulence: application to rotating turbulence}.
  \jt{Journal of Fluid Mechanics}  \bvol{908},  \pg{A17}.

\bibitem[Yousif {\em et~al.\/}(2022)Yousif, Yu, Hoyas, Vinuesa \&
  Lim]{yousif2022deep}
{\sc \au{Yousif, Mustafa~Z}, \au{Yu, Linqi}, \au{Hoyas, Sergio}, \au{Vinuesa,
  Ricardo} \& \au{Lim, HeeChang}} \yr{2022}  \at{A deep-learning approach for
  reconstructing 3d turbulent flows from 2d observation data}.  \jt{arXiv
  preprint arXiv:2208.05754} .

\bibitem[Zhang {\em et~al.\/}(2018)Zhang, Yuan, Zeng, Li \&
  Wei]{zhang2018missing}
{\sc \au{Zhang, Qiang}, \au{Yuan, Qiangqiang}, \au{Zeng, Chao}, \au{Li,
  Xinghua} \& \au{Wei, Yancong}} \yr{2018}  \at{Missing data reconstruction in
  remote sensing image with a unified spatial--temporal--spectral deep
  convolutional neural network}.  \jt{IEEE Transactions on Geoscience and
  Remote Sensing}  \bvol{56}~(8),  \pg{4274--4288}.

\end{thebibliography}

\end{document}